\documentclass[preprint2]{aastex6}

\usepackage{url}
\shorttitle{Mass Loading Efficiency and Particle Acceleration Efficiency of Relativistic Jets}
\shortauthors{Inoue \& Tanaka}

\begin{document}
\title{Baryon Loading Efficiency and Particle Acceleration Efficiency of Relativistic Jets: Cases For Low Luminosity BL Lacs} 

\author{Yoshiyuki Inoue\altaffilmark{1,2}} 
\affil{Institute of Space and Astronautical Science JAXA, 3-1-1 Yoshinodai, Chuo-ku, Sagamihara, Kanagawa 252-5210, Japan}

\author{Yasuyuki T. Tanaka} 
\affil{Hiroshima Astrophysical Science Center, Hiroshima University, 1-3-1 Kagamiyama, Higashi-Hiroshima, Hiroshima 739-8526, Japan}

\altaffiltext{1}{JAXA International Top Young Fellow}
\altaffiltext{2}{yinoue@astro.isas.jaxa.jp}

\begin{abstract}
Relativistic jets launched by supermassive black holes, so-called as active galactic nuclei (AGNs), are known as the most energetic particle accelerators in the universe. However, the baryon loading efficiency onto the jets from the accretion flows and their particle acceleration efficiencies have been veiled in mystery. With the latest data sets, we perform multi-wavelength spectral analysis of quiescent spectra of 13 TeV gamma-ray detected high-frequency-peaked BL Lacs (HBLs) following one-zone static synchrotron-self-Compton (SSC) model. We determine the minimum, cooling break, and maximum electron Lorentz factors following the diffusive shock acceleration (DSA) theory. We find that HBLs have $P_B/P_e\sim6.3\times10^{-3}$ and the radiative efficiency $\epsilon_{\rm rad,jet}\sim6.7\times10^{-4}$ where $P_B$ and $P_e$ is the Poynting and electron power, respectively. By assuming 10 leptons per one proton, the jet power relates to the black hole mass as $P_{\rm jet}/L_{\rm Edd}\sim0.18$ where $P_{\rm jet}$ and $L_{\rm Edd}$ is the jet power and the Eddington luminosity, respectively. Under our model assumptions, we further find that HBLs have the jet production efficiency of $\eta_{\rm jet}\sim1.5$ and the mass loading efficiency of $\xi_{\rm jet}\gtrsim5\times10^{-2}$. We also investigate the particle acceleration efficiency in the blazar zone by including the most recent {\it Swift}/BAT data. Our samples ubiquitously have the particle acceleration efficiency of $\eta_g\sim10^{4.5}$, which is inefficient to accelerate particles up to the ultra-high-energy-cosmic-ray (UHECR) regime. This implies that the UHECR acceleration sites should be other than the blazar zones of quiescent low power AGN jets, if one assumes the one-zone SSC model based on the DSA theory.
\end{abstract}


\section{Introduction}
\label{intro}

Baryon mass loading and particle acceleration efficiencies of relativistic jets of active galactic nuclei (AGNs) powered by supermassive black holes (SMBHs) have been major problems in astrophysics. The former is essential for understanding of the jet launching mechanism \citep[see e.g.][]{mck12,tom12} and the latter is for understanding of the origin of ultra-high-energy cosmic rays (UHECRs) whose energies are greater than $10^{18.5}$~eV \citep[see e.g.][for reviews]{kot11}. 

First, a relativistic outflow from a SMBH is powered by releasing its gravitational potential energy implying $P_{\rm jet}\lesssim\dot{M}_{\rm acc} c^2$, where $P_{\rm jet}$ is the total jet power, $\dot{M}_{\rm acc}$ is the mass accretion rate from the disk to the SMBH, and $c$ is the speed of light. Note that recent systematic analysis for luminous blazars by \citet{ghi14} shows the jet power is slightly larger than the power of accreting plasma by extracting the rotational energy of central SMBHs ($P_{\rm jet} \sim 1.4 \dot{M}_{\rm acc} c^2$). The jet power is made of the Poynting power $P_B$ and kinetic power $P_k=\Gamma \dot{M}_{\rm jet} c^2$ where $\Gamma$ is the bulk Lorentz factor and $\dot{M}_{\rm jet}$ is the mass outflow rate in the jet. As AGN jets typically have $\Gamma\sim10$--100, the mass loading efficiency $\xi_{\rm jet}$ is generally expected to be $\xi_{\rm jet}\equiv\dot{M}_{\rm jet}/\dot{M}_{\rm acc}<1$.

Systematic spectral studies of luminous blazars revealed that powerful relativistic jets generated by standard disk accretion \citep{ghi11} have the mass loading efficiency of an order of unity as $\xi_{\rm jet}\sim0.1-1$ \citep{cel08,ghi14}. On the contrary, the mass loading efficiency of low power jets which are supposed to be supported by the radiative-inefficient accretion flow (RIAF) disks is not well understood from observations, while recent 3D global general relativistic magetohydrodynamic (GRMHD) simulations for the jet launching from a RIAF disk indicates $\xi_{\rm jet}\sim0$-0.2 depending on parameters such as the SMBH spin \citep{mck12}.

Second, the most energetic particle accelerators in the universe are the AGN jets. Although the AGN jets have been believed as the promising sites to accelerate the UHECRs \citep[e.g.][]{bie87, tak90, rac93}, the particle acceleration efficiency in the jets have been veiled in mystery for a long time. Among jet dominated AGNs, low power jet populations, HBLs and Fanaroff Riley (FR)~I radio galaxies \citep{fan74}, are more likely to be the sources of UHECRs than powerful jet populations, flat spectrum radio quasars (FSRQs) and FR~II radio galaxies, under the assumption of the Bohm limit acceleration, as HBLs and FR~Is locate within the Greisen-Zatsepin-Kuzmin (GZK) horizon \citep{gre66,zat66} and have sufficient emissivity to power UHECRs \citep{der10}. Observationally, it is also known that arrival directions of UHECRs show an excess in the vicinity of the FR~I radio galaxy Centaurus~A \citep[e.g.][]{abr08,abr10_prl}.

Determination of the particle acceleration efficiency in the jets  is essential for the understanding of the origin of UHECRs. Early blazar spectral analysis studies have shown the acceleration efficiency can be as large as $\sim10^5$ by assuming the maximum electron energy is given by the balance between acceleration and cooling time scales \citep{ino96,sat08,fin08}. However, the number of samples was few and non-simultaneous data have been adopted. Therefore, detailed spectral analysis studies with large HBL samples are demanded to investigate the mass loading efficiency and the particle acceleration efficiency in the blazar zone of low power jet populations. 

Here, multi-wavelength electromagnetic observations from radio to gamma-ray allow us to study overall spectral energy distribution (SED) and physical parameters of AGN jets. The observed blazar SEDs consist of two non-thermal broadband components \citep[e.g.][]{tak96,abd10_blz} while additional thermal disk components appear from infrared to X-ray in the case of radio galaxies \citep[e.g.][]{kat11} which are a misaligned population of blazars \citep{urr95}. The low-energy non-thermal component is electron synchrotron radiation and the other non-thermal component is inverse Compton component in which electrons scatter internal synchrotron radiation field, so-called synchrotron self-Compton (SSC) radiation \citep[e.g.,][]{jon74,mar92,tav98,fin08} or external radiation field, so-called external Compton (EC) radiation \citep[e.g.][]{der93,sik94}. Peak energies of synchrotron and inverse Compton components are known to decrease with increasing bolometric luminosities \citep{fos98,kub98}. This is called as the blazar sequence, although the validity of the sequence is still under debate due to possible selection biases \citep{pad07,gio12}.

The overall non-thermal spectra of HBLs and FR~I galaxies are well fit with the leptonic SSC scenario \citep[e.g.][]{tav98,abd09_ngc1275}. Those models allow us to estimate the energetics of jets such as Poynting, kinetic, and total jet powers \citep[e.g.][]{tav98,cel08,tav10}. The SSC scenarios can also reproduce the correlated flux variabilities in X-ray and TeV gamma-ray bands \citep[e.g.][]{mat97,tak00,li00}. However, it would be challenging for the SSC models to explain orphan gamma-ray flares \citep[e.g.][]{kra04} and minutes-scale time scale variability \citep{aha07_flare,alb07_Mrk501}. Considering inhomogeneous emitting regions, the SSC scenario can also explain the orphan flares \citep{kus06}. 

Alternatively, hadronic or leptohadronic scenarios have also been considered as the gamma-ray emission mechanism of blazars \citep[e.g.][]{aha00,muc01}. Relativistic hadrons may explain the high energy emission from radio-loud AGNs through proton-synchrotron processes or cascade emission from photohadronic interactions. However, hadronic processes are known to be inefficient and requires super-Eddington jet powers as $P_{\rm jet}\sim100 L_{\rm Edd}$ where $L_{\rm Edd}$ is the Eddington luminosity \citep{sik09,zdz15}. 

In each scenario, particles need to be accelerated to high energies at relativistic shock fronts to radiate high energy photons. The most promising acceleration mechanism is the diffusive shock acceleration (DSA), so-called first-order Fermi acceleration, \citep[e.g.][]{kry77,axf78,bel78,bla78}. Under suitable conditions, shocks can efficiently accelerate high energy particles \citep[e.g.][]{dru83,bla87,jon91}. However, its particle acceleration efficiency, the Bohm diffusion coefficient $\eta_g$ in the jets has been debated for a long time. To achieve UHECR energies in AGN jets, it requires near the Bohm limit $\eta_g=1$ \citep[e.g.][]{der10}.

In this paper, we investigate the energetics and the particle acceleration efficiency in the blazar zone of HBLs using the one-zone static SSC model in the context of the DSA theory. Although stochastic acceleration, so-called the second-order Fermi acceleration \citep[e.g.][]{yan13,asa14}, and reconnection acceleration \citep[e.g.][]{sir15} are also recently discussed as an acceleration mechanism in the blazar zone, we focus on the first-order Fermi acceleration scenario in this paper. 

Multi-wavelength SED data in the quiescent state is adopted in this paper. Various observatories are continuously monitoring the high energy sky. Currently, Monitor of All-sky X-ray Image ({\it MAXI})/Gas Slit Camera (GSC) observes the sky at 0.5--30~keV \citep{mat09}, the {\it Swift}/Burst Alert Telescope (BAT) at 15--150~keV \citep{bar05,kri13}, Astrorivelatore Gamma ad Immagini LEggero ({\it AGILE}) at 30~MeV-50~GeV \citep{tav09}, and the large area telescope (LAT) onboard the {\it Fermi} gamma-ray space telescopes (hereinafter {\it Fermi}) at 20~MeV--300~GeV \citep{atw09}. These monitoring data can provide the SEDs of AGNs. And, ground based imaging atmospheric Cherenkov telescopes (IACTs) have observed $\sim50$ blazars in various states \citep{tevcat}\footnote{\url{ http://tevcat.uchicago.edu/}}.

This paper is organized as follows. In Section \ref{sec:sam}, we introduce the sample used in our analysis which are detected in X-ray and gamma-ray. In Section \ref{sec:model}, we present the spectral modelling and the fitting method. In Section \ref{sec:para}, fitting results and the inferred parameters are presented. Jet energetics and particle acceleration are discussed in Section \ref{sec:power} and  Section \ref{sec:PAE}, respectively. Discussion and conclusion is given in Section \ref{sec:dis} and Section \ref{sec:sum}, respectively. Throughout this paper, we adopt the standard cosmological parameters of $(h, \Omega_M , \Omega_\Lambda) = (0.7, 0.3, 0.7)$.

\section{Samples}
\label{sec:sam}

\floattable
\begin{deluxetable*}{lcccccc}
\rotate
 \tablecaption{HBL samples\label{tab:sample}}
\tablehead{
  \colhead{Source} & \colhead{Redshift}  & \colhead{BH mass [$\log M_\odot$]} & \colhead{{\it MAXI}/GSC Name} & \colhead{{\it Swift}/BAT Name} & \colhead{{\it Fermi}/LAT Name} & \colhead{IACTs}\
 }
\startdata
Mrk~421 & 0.031 & 8.3\tablenotemark{a} & 2MAXI~J1104+382 & SWIFT~J1104.4+3812 & 3FGL~J1104.4+3812  &  MAGIC \citep{alb07_mrk421} \\
Mrk~501 & 0.034 & 9.2\tablenotemark{a} & 2MAXI~J1653+398  & SWIFT~J1654.0+3946 & 3FGL~J1653.9+3945  &  MAGIC \citep{alb07_Mrk501}\\
1ES~2344+514& 0.044 & 8.8\tablenotemark{a} & - & SWIFT~J2346.8+5143 & 3FGL~J2347.0+5142  & MAGIC \citep{magic07_1ES2344+514}\\ 
1ES~1959+650 & 0.048 & 8.1\tablenotemark{a} & 2MAXI~J1959+651 & SWIFT~J1959.6+6507 & 3FGL~J2000.0+6509  & MAGIC \citep{magic06_1ES1959+650}\\
PKS~0548-322  & 0.069 & 8.2\tablenotemark{a} & - & SWIFT~J0550.7-3212A& 3FGL~J0550.6-3217  & H.E.S.S. \citep{hess10_PKS0548-322}\\
PKS~2005-489  & 0.071 & 9.0\tablenotemark{b} & 2MAXI~J2009-487 & SWIFT~J2009.6-4851 & 3FGL~J2009.3-4849  & H.E.S.S. \citep{hess10_PKS2005-489}\\
RGB~J0710+591  & 0.125 & 8.3\tablenotemark{a} & 2MAXI~J0710+592 & SWIFT~J0710.3+5908 & 3FGL~J0710.3+5908  & VERITAS \citep{acc10_RGBJ0710+591}\\
H~1426+428  & 0.129 & 9.1\tablenotemark{b} & 2MAXI~J1429+425 & SWIFT~J1428.7+4234 & 3FGL~J1428.5+4240  & H.E.S.S. \citep{aha02_H1426+428}\\
1ES~0229+200  & 0.14 & 9.2\tablenotemark{b} & - & SWIFT~J0232.8+2020 & 3FGL~J0232.8+2016  & H.E.S.S. \citep{aha07_1ES0229+200}\\
H~2356-309  & 0.165 & 8.6\tablenotemark{b} & 2MAXI~J2359-307 & SWIFT~J2359.0-3038 & 3FGL~J2359.3-3038  & H.E.S.S. \citep{aha06_H2356-309}\\
1ES~1218+304  & 0.182 & 8.6\tablenotemark{b} & - & SWIFT~J1221.3+3012 & 3FGL~J1221.3+3010  &  VERITAS \citep{acc09_1ES1218+304}\\
1ES~1101-232  & 0.186 & 9.0\tablenotemark{c} & 2MAXI~J1104+382 & SWIFT~J1103.5-2329 & 3FGL~J1103.5-2329  & H.E.S.S. \citep{aha07_1ES1101-232}\\
1ES~0347-121  & 0.188 & 8.7\tablenotemark{b} & 2MAXI~J0348-120 & SWIFT~J0349.2-1159 & 3FGL~J0349.2-1158 &  H.E.S.S.\citep{aha07_1ES0347-121}\\
\enddata
\tablecomments{Redshift values are taken from the TeVcat data \citep{tevcat}. The reference for {\it MAXI}/GSC, {\it Swift}/BAT, and {\it Fermi}/LAT is \citet{hir13}, \citet{bau13}, and \citet{ace15_3fgl}, respectively. The references for the TeV gamma-ray data are shown in the column of IACTs in the table. The references for the BH mass estimation are \citet{woo02,woo05}.}
\tablenotetext{a}{The mass is estimated by using the correlation between $M_{\rm BH}$ and the stellar velocity dispersion.}
\tablenotetext{b}{The mass is estimated by using the fundamental plane among the stellar velocity dispersion, the surface brightness, and the effective radii.}
\tablenotetext{c}{No measurements of the central black hole mass of 1ES~1101-232 are available. The fiducial value of $M_{\rm BH}=10^9 M_{\odot}$ is adopted here.}
\end{deluxetable*}

Blazars are divided into two categories by their optical spectral features. Those are BL~Lacertae objects (BL~Lacs) and FSRQs \footnote{FSRQs are also called as quasar-hosted blazars (QHBs).}. BL~Lacs have the emission lines with the equivalent width of $\lesssim5$~\AA. BL~Lacs are further divided by their synchrotron spectral peak positions $\nu_{\rm syn}$ into low-frequency peaked BL Lacs (LBLs; $\nu_{\rm syn}\le10^{14.1}$~Hz), intermediate-frequency peaked BL Lacs (IBLs; $10^{14.1}~{\rm Hz}\le\nu_{\rm syn}<10^{14.8}$~Hz), and high-frequency peaked BL Lacs (HBLs; $10^{14.8}~{\rm Hz}\le \nu_{\rm syn}$)\footnote{There are alternative classification as low-synchrotron-peaked blazars (LSPs; $\nu_{\rm syn}<10^{14}$~Hz), intermediate-synchrotron-peaked blazars (ISP; $10^{14}~{\rm Hz}<\nu_{\rm syn}<10^{15}$~Hz) and high-synchrotron-peaked blazars (HSP; $10^{15}~{\rm Hz}< \nu_{\rm syn}$). See \citet{abd10_agn} for details.}. HBLs are known to extend their emission up to the TeV band \citep[e.g. Mrk~421;][]{abd11_mrk421}. Following the blazar sequence, HBLs tends to be less luminous blazar populations.

To study the multi-wavelength properties of HBLs, we select HBL samples from the default catalog of TeVcat \citep{tevcat}, which includes only sources reported in refereed journals. There are 33 HBLs, of which 29 HBLs have redshift information\footnote{We do not include PG~1553+113 in our sample, although the redshift of the source has recently been constrained in the range of $z=0.49\pm0.04$ \citep{abr15_pg1553}. }. Among them we select sources whose low-state spectrum data are available \citep[see][for details]{ino16} and which are reported in the {\it Swift}/BAT 70-months catalog \citep{bau13}. After these selection cuts, there are 13~HBLs as listed in Table \ref{tab:sample}. 

As blazars are known to be variable sources, we focus on the quiescent state to combine multi-wavelength all sky monitoring observations. Since X-ray data is crucial for this study to determine the particle acceleration efficiency, we select the sources having the {\it Swift}/BAT data. To obtain the {\it Swift}/BAT hard X-ray spectral data points, we analyze the {\it Swift}/BAT 70-month survey data using the pha data files and the response matrix provided at \url{http://swift.gsfc.nasa.gov/results/bs70mon/} \citep{bau13}. We also obtain the {\it Fermi}/LAT data from the 3FGL catalog for GeV gamma-ray data \citep{ace15_3fgl}\footnote{\url{http://fermi.gsfc.nasa.gov/ssc/data/access/lat/4yr_catalog/}} and the {\it MAXI}/GSC data from the 37-month MAXI catalog \citep{hir13}.  All of our HBL samples have the 3FGL counterparts. The redshift information is taken from the TeVcat database. The corresponding object identification names and references for TeV data are listed in Table \ref{tab:sample}. Although the {\it Fermi} source catalog detected above 50 GeV \citep[2FHL;][]{ack15_2fhl} has recently been released, we do not include those data.

We further include the central SMBH mass $M_{\rm BH}$ information of our samples. The mass of the nuclear black holes provide fundamental information for the jet study such as the Schwarzschild radius and the Eddington luminosity. Since spatially resolved kinematics observations are limited to only nearby sources, various indirect methods have been developed such as the reverberation mapping \citep[e.g.][]{bla82,pet93,kas00}, the correlation between the optical luminosity and the broad-line-region size \citep[e.g.][]{kas00,mcl01,ves02}, the correlation between $M_{\rm BH}$ and the stellar velocity dispersion \citep[e.g.][]{fer00,geb00,tre02}, and the fundamental plane among the stellar velocity dispersion, the surface brightness, and the effective radii \citep[e.g.][]{jor96,woo02,woo05}. 

In our paper, we use the value estimated in \citet{woo02} and \citet{woo05} (Table \ref{tab:sample}). $M_{\rm BH}$ is estimated by the stellar velocity dispersion method for Mrk~421, Mrk~501, 1ES~2344+514, 1ES~1959+590, PKS~0548-322, and RGB~J0710+591, and by the fundamental plane method for PKS~2005-489, H~1426+428, 1ES~0229+200, H~2356-309, 1ES~1218+304, and 1ES~0347-121. As mass measurements of the central SMBH of 1ES~1101-232 are not available, the fiducial value of $M_{\rm BH}=10^9 M_{\odot}$ is adopted in this paper. The average mass of the central SMBHs in our sample is $<M_{\rm BH}>\sim5\times10^8M_{\odot}$.

\section{Spectral Energy Distribution Modelling}
\label{sec:model}
We consider a spherical emitting plasma located at distance $r$ from the central SMBH moving with velocity $\beta=v/c$ and Lorentz factor $\Gamma=(1-\beta^2)^{-1/2}$ where $c$ is the speed of light. The Doppler beaming factor is given by $\delta=[\Gamma(1-\beta\cos\theta_{\rm obs})]^{-1}$ for an observer located at a viewing angle $\theta_{\rm obs}$ with respect to the plasma velocity vector. $\delta$ and $\Gamma$ is expected to have the same order in blazars. For the simplicity, we adopt $\delta=\Gamma$. 

We set the comoving radius of the spherical emitting region $R'$ which is related to the comoving variability timescale $t_{\rm var}'$ as $R'\lesssim ct_{\rm var}'$. Quantities in the comoving frame of the emitting region are primed hereinafter. For the observer, the measured variability time scale is $t_{\rm var}=(1+z)t_{\rm var}'/\delta$ where $z$ is the redshift of the source. In this paper, however, we relate $R'$ to the location of the emitting plasma via $R'\simeq r\theta$ where $\theta$ is the angle size of the emitting blob with respect to the central SMBH. Although it is not clear whether $\theta$ is equal to the jet opening angle $\theta_j$, we assume $\theta=\theta_j$ for the convenience. Since recent numerical simulations of relativistic jets confined by external pressure find $\Gamma\theta_j\lesssim1$ \citep[see e.g.][]{kom09}, we set $\Gamma\theta_j=1$ thus $\theta=1/\delta$.

The location of the emitting plasma of blazars is not well constrained. For the simplicity, we assume $r=3000 r_s$ in our analysis, where $r_s=2GM_{\rm BH}/c^2$ is the Schwarzschild radius where $G$ is the gravitational constant. $r=3000r_s$ roughly corresponds to $t_{\rm var}\sim 1$~day for $\delta=10$ and $M_{\rm BH}=5\times10^8M_\odot$ which is the typical variability time scale of blazars. We consider different emission locations in Section \ref{subsec:diff_r}.

To get the electron distribution, we need to solve the kinetic equation in the steady state. For the simplicity, however, we assume that the electron distribution is given by the following form
\begin{eqnarray}
\nonumber
\frac{dN_e'}{d\gamma'}(\gamma')&=& K_e  \gamma_b'^{-p_1} \left[\left(\frac{\gamma'}{\gamma_{b}'}\right)^{p_1} + \left(\frac{\gamma'}{\gamma_{b}'}\right)^{p_2} \right]^{-1}\\
&\times&\exp{\left[-\left(\frac{\gamma'}{\gamma_c}\right)^{\alpha}\right]}H(\gamma'-\gamma'_{\rm min}),
\label{eq:Ne}
\end{eqnarray} 
where $\gamma'$ is the Lorentz factor of an electron in the jet comoving frame, $\gamma_b'$ is the cooling break electron Lorentz factor determined from the balance between cooling energy loss and dynamical time scales, $\gamma_c'$ is the maximum electron Lorentz factor determined from the balance of the cooling loss and acceleration, and $\gamma'_{\rm min}$ is the minimum electron Lorentz factor (See Eq.  \ref{eq:gmin}). $H(x)$ is the Heaviside function defined as $H(x)=0$ for $x<0$ and $H(x)=1$ for $x\ge0$. If $\gamma_c'<\gamma_b'$, we omit the term of $(\gamma'/\gamma_b')^{p_2}$ in Equation \ref{eq:Ne}. 

The cooling break index must be $\Delta p\equiv p_2-p_1=1$ \citep[e.g.][]{lon94}. This leads the spectral steepening of 0.5 for synchrotron and inverse-Compton emissions in the homogeneous stationary fluid. However, for inhomogeneous sources, different amounts of radiation spectral steepening can be achieved depending on their fluid velocity, geometrical structure, magnetic field structure, and density distribution \citep{rey09}. Such inhomogeneities will enables SSC emission models to explain the orphan flares \citep{kus06}. As the inhomogeneity of the emitting blob is hard to treat in our modelling, we phenomenologically include the effect of inhomogeneity in $p_2$ by setting it as a free parameter. As pointed in \citet{abd11_mrk421, abd11_mrk501}, the electron spectrum breaks for some sources are needed to be different from unity. 

Electrons lose their energies via synchrotron and inverse-Compton radiation. The synchrotron cooling rate in the comoving frame is
\begin{equation}
\dot{\gamma}'_{\rm syn}(\gamma') = -\frac{4c\sigma_TU_B\gamma'^2}{3m_ec^2},
\end{equation}
where $m_e$ is the electron rest mass, $\sigma_T$ is the Thomson cross section and $U_B =B^2/8\pi$ is the magnetic field energy density of magnetic field strength $B$. The SSC cooling rate including the Klein-Nishina cross section \citep{jon68,boe97,fin08,der09} is
\begin{equation}
\dot{\gamma}'_{\rm SSC}(\gamma') = -\frac{3\sigma_T}{8m_ec}\int_0^{\infty}d\epsilon'\frac{u_{\rm syn}'(\epsilon')}{\epsilon'^2}G(\gamma'\epsilon'),
\end{equation}
where $\epsilon'=E_\gamma'/m_ec^2$, $E_\gamma'$ is the photon energy, $u_{\rm syn}'(\epsilon')$ is the synchrotron energy density, and
\begin{eqnarray}
G(E) &=& \frac{8E(1+5E)}{3(1+4E)^2}-\frac{4E}{1+4E}\left(\frac{2}{3}+\frac{1}{2E}+\frac{1}{8E^2}\right) \nonumber \\
+\ln(&1&+4E)\left[1+\frac{3}{E}+\frac{3}{4E^2}+\frac{\ln(1+4E)}{2E}-\frac{\ln(4E)}{E}\right]\nonumber\\
-\frac{5}{2E}&+&\frac{1}{E}\sum_{n=1}^{\infty}\frac{(1+4E)^{-n}}{n^2}-\frac{\pi^2}{6E}-2.
\end{eqnarray}
The total cooling rate is given by $\dot{\gamma}'_{\rm cool}(\gamma')=\dot{\gamma}'_{\rm syn}(\gamma')+\dot{\gamma}'_{\rm SSC}(\gamma')$
and the cooling time scale is 
\begin{equation}
\label{eq:cool}
t'_{\rm cool}(\gamma')=\frac{\gamma'}{|\dot{\gamma}'_{\rm syn}(\gamma')+\dot{\gamma}'_{\rm SSC}(\gamma'))|}.
\end{equation} 

The dynamical time scale is estimated as 
\begin{equation}
\label{eq:dyn}
t'_{\rm dyn}(\gamma')=\frac{R'}{c}.
\end{equation}

For the acceleration time scale $t_{\rm acc}'$, we assume electrons are accelerated by DSA. In the frame work of DSA \citep[e.g.,][]{dru83,bla87}, $t_{\rm acc}'$ can be approximated as 
\begin{equation}
\label{eq:acc}
t_{\rm acc}'(\gamma')=\frac{\eta_{\rm acc}D'(\gamma')}{u_{\rm sh}'^2},
\end{equation}
where $D'({\gamma'})$ is the diffusion coefficient and $u_{\rm sh}'=\beta c\simeq c$ is the shock speed. We set $\eta_{\rm acc}=10$. Assuming the Bohm diffusion, 
\begin{equation}
\label{eq:diff}
D'(\gamma')=\frac{\eta_gc\gamma'm_ec^2}{3eB},
\end{equation}
where $e$ is the electric charge and $\eta_g$ is the gyrofactor. It becomes the Bohm limit with $\eta_g=1$. 

In the non-relativistic shocks, the spectral shape factor $\alpha$ (See Eq. \ref{eq:Ne}) is related to the index which represents the energy dependence of the diffusion coefficient $\beta_B$ ($D'(\gamma')\propto\gamma'^{\beta_B}$) as $\alpha = \beta_B+1$ in the cooling limited case \citep[e.g.,][for non-relativistic shocks]{zir07,yam14,yam15}. As we assume the Bohm diffusion ($\beta_B=1$), $\alpha=2$. If the energy dependence of diffusion is weaker, $\alpha$ becomes smaller. $\alpha$ is also $\sim2$ for relativistic shocks based on a semi-analytical model \citep{dem07}. 

If the cooling is dominated by synchrotron emission, we approximately have 
\begin{eqnarray}
\gamma_b' &=& \frac{6\pi m_ec^2}{\sigma_T B^{2}R'}\simeq 7.7\times10^4 \left(\frac{B}{0.1~{\rm G}}\right)^{-2}\left(\frac{R'}{3\times10^{16}~{\rm cm}}\right)^{-1},\label{eq:gb} \\  
\gamma_c'&=&\left(\frac{18\pi e}{\sigma_T\eta_{\rm acc}\eta_gB}\right)^{1/2}\simeq 2.0\times10^6 \left(\frac{\eta_g}{10^4}\right)^{-1/2}\left(\frac{B}{0.1~{\rm G}}\right)^{-1/2}. \label{eq:gc}
\end{eqnarray}
Because of the Klein-Nishina effect, synchrotron radiation dominates the cooling effect at $\sim\gamma_c'$ even if we include cooling effect by SSC emission. Moreover, the spectral break shape would not be a pure power-law break due to the transition from the Thomson regime to the Klein-Nishina regime.

We also consider the dynamical time-limited case in which the maximum energy is limited by the dynamical time scale $t_{\rm acc}'=t_{\rm dyn}'<t_{\rm cool}'$. This leads
\begin{equation}
\gamma_c' = \frac{3eBR'}{\eta_{\rm acc}\eta_gm_ec^2} \simeq 5.3\times10^{7} \left(\frac{B}{0.1~{\rm G}}\right)\left(\frac{R'}{3\times10^{16}~{\rm cm}}\right)\left(\frac{\eta_g}{10^4}\right)^{-1}.
\end{equation} 
In the dynamical time-limited case, $\alpha=2 \beta_B$ \citep[e.g.,][]{yam15}. As we follow the Bohm diffusion ($\beta_B=1$), $\alpha$ is the same as the cooling limited case. 

\begin{figure*}
 \begin{center}
  \includegraphics[width=16cm]{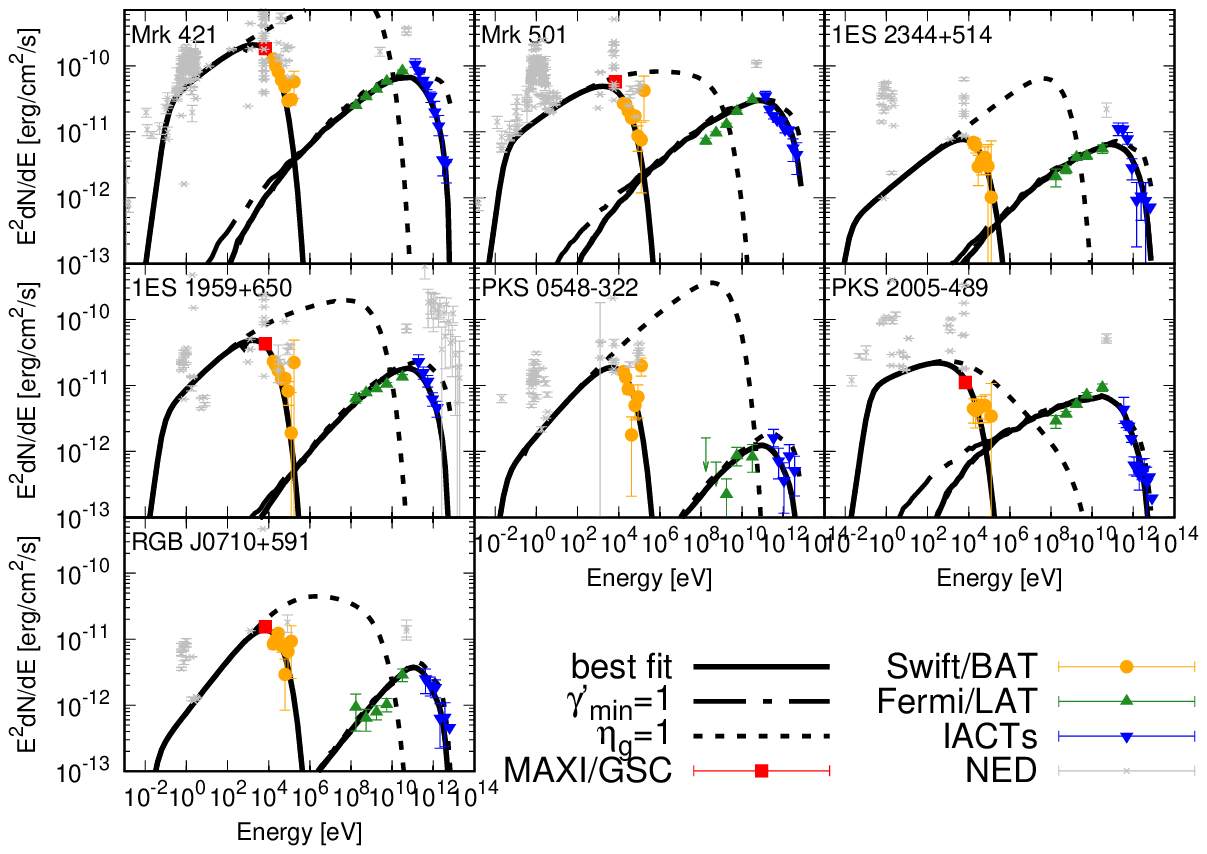} 
 \end{center}
\caption{Mutli-wavelength spectral energy distribution of the HBLs in the low state. The 37-month averaged {\it MAXI}/GSC \citep{hir13}, 70-month averaged {\it Swift}/BAT \citep{bau13}, 48-month averaged {\it Fermi}/LAT \citep{ace15_3fgl}, and quiescent state IACTs data (see Table \ref{tab:sample} for the references information) are shown by red square, orange circle, green top-triangle, and blue down-triangle data points, respectively. Archival data from NED are also shown. The solid curve shows the best-fit model.  The dot-dashed and dot curve represents the model with $\gamma'_{\rm min}=1$ and $\eta_g=1$, respectively, but the other parameters are unchanged from the best-fit model. The source name is indicated in each panel. See the text for more details.}\label{fig:SED_v1}
\end{figure*}

The acceleration of electrons becomes efficient in collisionless shocks, if they are freely able to cross the shock front. The condition for the acceleration depends on the plasma content. In this paper, we assume that the plasma is composed of $q$ leptons per one proton as
\begin{equation} 
q \equiv \frac{N_{e^-} + N_{e^+}}{N_p},
\end{equation}
where $N_{e^-}$, $N_{e^+}$, and $N_p$ is the total electron, positron, and proton number in the blob, respectively. 

Pure pair plasma jet models are excluded from X-ray observations \citep{sik00} and pairs may not survive the annihilation in the inner, compact, and dense regions \citep{cel08,ghi10}, although there is still room for pairs in the jet based on the energetics arguments~\citep{sik00,sik05}. Based on X-ray and gamma-ray observations, \citet{sik05} suggested that $N_{e^-}/N_p\simeq20(\eta_{\rm diss}\eta_e/0.1)/(\bar{\gamma}_{e, \rm inj}/10)$ leading $q\sim10$ where $\eta_{\rm diss}$ is the efficiency of the energy dissipation in the blazar zone, $\eta_e$ is the fraction of the dissipated energy converted to relativistic electrons, and $\bar{\gamma}_{e, \rm inj}$ is the average Lorentz factor of relativistic electrons. \citet{kat08_1510} reported the existence of the soft X-ray excess in the spectrum of the FSRQ PKS~1510-089 which can be explained as the bulk Compton features interpreting $N_{e^-}/N_p\sim10$. Alternatively, this excess can be also explained as a contribution of the SSC component or the non-blazar AGN disk emission component \citep{kat08_1510}. We take $q=10$ as the fiducial values, otherwise noted. We test other $q$ values in Sec. \ref{subsec:diff_q}.

The requirement for the minimum electron energy to be operated by the shock acceleration is determined as \citep[see e.g.][]{pir99,der09}
\begin{equation}
\label{eq:gmin}
\frac{\int_{\gamma'_{\rm min}}^{\infty}\gamma'\frac{dN_e'}{d\gamma'}(\gamma')d\gamma'}{\int_{\gamma'_{\rm min}}^{\infty}\frac{dN_e'}{d\gamma'}(\gamma')d\gamma'}\simeq \frac{\epsilon_e}{q} \frac{m_p}{m_e}\Gamma_{\rm sh},
\end{equation}
where $\epsilon_e$ represents the fraction of which the shock energy is transferred to the acceleration of leptons and $\Gamma_{\rm sh}$ represents the typical relative Lorentz factor between internal shock shells. We assume $\Gamma_{\rm sh}=2$. If $p_1>2$, $\gamma'_{\rm min} \simeq \epsilon_e\Gamma_{\rm sh} m_p(p_1-2)/qm_e(p_1-1)$. We adopt $\epsilon_e=0.1$ as the fiducial value in this paper, i.e. 10\% of shock energy goes into electron acceleration. Multi-wavelength spectral fits for Mrk~501 suggest $\epsilon_e\sim0.1$ \citep{abd11_mrk501}. Number of electrons at $1\le\gamma'<\gamma'_{\rm min}$ should be sufficiently small comparing to entire electron numbers.

In this paper, we determine $\gamma_b'$ and $\gamma_c'$ by balancing cooling (Equation \ref{eq:cool}) and dynamical (Equation \ref{eq:dyn}) time scales  and cooling (Equation \ref{eq:cool}) and acceleration (Equation \ref{eq:acc}) time scales, respectively. And, we find $\gamma'_{\rm min}$ by solving Equation \ref{eq:gmin}, otherwise we notice. Therefore, we self-consistently derive $\gamma'_{\rm min}$, $\gamma'_b$, and $\gamma'_c$ in the frame work of the DSA theory once we determine other parameters.

\begin{figure*}
 \begin{center}
  \includegraphics[width=16cm]{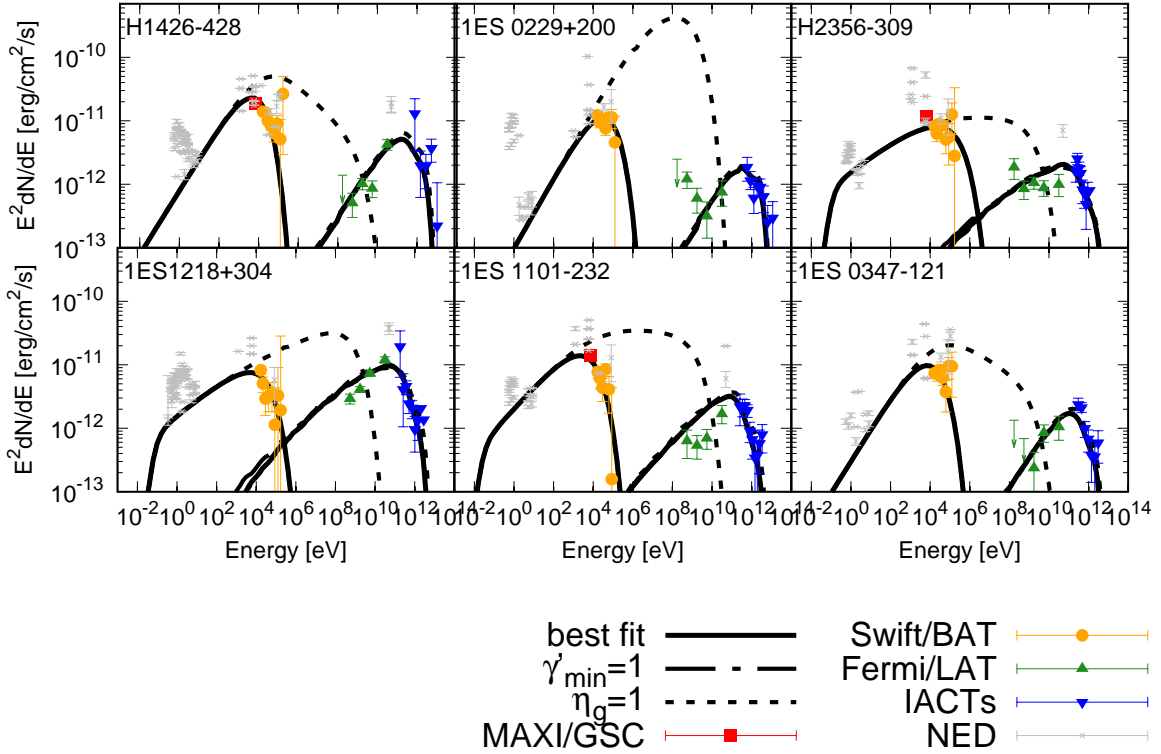} 
 \end{center}
\caption{Same as \autoref{fig:SED_v1}, but for the other HBL samples as indicated in each panel.}\label{fig:SED_v2}
\end{figure*}

For the calculation of synchrotron and SSC emission components, we follow the exact expression of synchrotron emission and SSC emission taking into account the Klein-Nishina cross section following \citet{fin08}. We include synchrotron self-absorption (SSA) following Equation 7.145 of \citet{der09}\footnote{The original equation of \citet{der09} had wrong sign \citep[see][]{der14}.} with the $\delta$-function approximation in which the SSA coefficient is
\begin{equation}
\kappa_{\nu'} = \frac{-\pi c r_e}{36\nu'}\left[\gamma'^2\frac{\partial}{\partial\gamma'}\left(\frac{n_e'(\gamma')}{\gamma'}\right)\right],
\end{equation}
where $n_e'(\gamma')$ is the electron density and $\gamma'=(\nu'/\nu_B)^{1/2}$. $\nu_B=m_ec^2B/hB_{\rm cr}$. $h$ is the Planck constant and $B_{\rm cr}=4.414\times10^{13}$~G is the critical magnetic field. The SSA opacity is $\tau_{\rm SSA}=2\kappa R'$. The absorbed spectrum is given by multiplying the factor $3u(\tau_{\rm SSA})/\tau_{\rm SSA}$, where $u(\tau_{\rm SSA})=1/2+\exp(-\tau_{\rm SSA})/\tau_{\rm SSA}-[1-\exp(-\tau_{\rm SSA})]/\tau_{\rm SSA}^2$ \citep{gou79, der09}. The SSA break in the comoving frame approximately appears at \citep[e.g.][]{nal14}
\begin{eqnarray}
\nu'_{\rm SSA} &\simeq& \frac{1}{3}\left(\frac{eB}{m_e^3c}\right)^{1/7} \frac{L'^{2/7}_{\rm syn}}{R'^{4/7}} \\
\simeq& 300 &\left(\frac{B}{0.1~{\rm G}}\right)^{\frac{1}{7}} \left(\frac{L'_{\rm syn}}{10^{43}~{\rm erg/s}}\right)^{\frac{2}{7}} \left(\frac{R'}{3\times10^{16}~{\rm cm}}\right)^{-\frac{4}{7}}~[{\rm GHz}] ,
\end{eqnarray}
where $L'_{\rm syn}$ is the synchrotron luminosity at the SSA break. Thus, the model can not account for the radio flux at $\lesssim300$~GHz ($\sim10^{-3}$~eV).

Gamma rays traveling the intergalactic space can be attenuated by photon-photon pair production interactions ($\gamma\gamma\rightarrow e^+e^-$) with low-energy photons of the extragalactic background light \citep[EBL; e.g.][]{gou66,jel66,ste92,fin10,dom11,ino13_cib}. For gamma rays of a given energy of $E_\gamma$, the pair production cross section peaks for low-energy photons with energy of $E\simeq2m_e^2c^4/E_\gamma\simeq0.5(1~{\rm TeV}/E_\gamma)$~eV. In this paper, we adopt the model by \citet{ino13_cib} which is based on a semi-analytical galaxy formation model including first stars. \citet{ino13_cib} is consistent with other EBL models \citep[see Figure 9 of][]{ino13_cib}.

The pairs generated by pair production subsequently up-scatter the cosmic microwave background (CMB) radiation to the GeV gamma-ray photons, so-called gamma-ray induced cascade emission \citep[e.g.][]{aha94,wan01,dai02,mur08}. Although plasma beam instability may suppress the cascade emission \citep[e.g.][]{bro12}, recent Particle-In-Cell simulations reveal that the plasma instability carries 10\% of the attenuated energy at most \citep{sir14}. Although the cascade gamma-ray spectra may affect the resulting spectra,  cascade processes are strongly affected by the intergalactic magnetic field strength and distribution \citep[e.g.][]{pla95,fin15} which are highly uncertain. Therefore, we do not include the cascade emission component in our analysis.

\floattable
\begin{deluxetable*}{lcccccc}
\tabletypesize{\footnotesize}
 \tablecaption{Parameter Fitting Results\label{tab:para}}
\tablehead{
  \colhead{Source} & \colhead{$\log K_e$}  & \colhead{$p_1$} & \colhead{$p_2$} & \colhead{$B$~[G]} & \colhead{$\delta$}& \colhead{$\log\eta_g$\tablenotemark{a}}\
 }
\startdata
Mrk~421		&	53.0	&	2.4	&	2.8			&	0.14	&	30	&	5.1$_{-0.0060}^{+0.0060}$\\
Mrk~501		&	55.8	&	2.5	&	3.1			&	0.011	&	17	&	4.6$_{-0.024}^{+0.018}$\\
1ES~2344+514	&	56.0	&	2.4	&	2.6			&	0.018	&	5.8	&	4.0$_{-0.090}^{+0.084}$\\
1ES~1959+650	&	52.2	&	2.3	&	2.9			&	0.14	&	30	&	5.1$_{-0.018}^{+0.012}$\\
PKS~0548-322	&	51.0	&	2.2	&	2.6			&	0.14	&	41	&	5.0$_{-0.036}^{+0.036}$\\
PKS~2005-489	&	55.4	&	2.7	&	3.7			&	0.030	&	30	&	5.1$_{-0.17}^{+0.13}$\\
RGB~J0710+591	&	51.6	&	2.2	&	-\tablenotemark{b}	&	0.050	&	47	&	4.5$_{-0.024}^{+0.018}$\\
H1426+428	&	53.0	&	2.0	&	-\tablenotemark{b}	&	0.011	&	22	&	4.7$_{-0.012}^{+0.012}$\\
1ES~0229+200	&	52.8	&	1.9	&	-\tablenotemark{b}	&	0.0013	&	26	&	3.0$_{-0.030}^{+0.030}$\\
H~2356-309	&	53.6	&	2.5	&	3.1			&	0.050	&	41	&	3.7$_{-0.42}^{+	0.24}$\\
1ES~1218+304	&	54.6	&	2.4	&	2.8			&	0.050	&	19	&	4.4$_{-0.11}^{+0.096}$\\
1ES~1101-232	&	53.4	&	2.3	&	3.1			&	0.030	&	41	&	5.1$_{-0.036}^{+0.024}$\\
1ES~0347-121	&	51.8	&	2.0	&	3.4			&	0.030	&	30	&	4.6$_{-0.036}^{+0.036}$\\
\enddata
\tablecomments{These parameters are set free in our fitting procedures. The other parameters derived from these fitting results are shown in Table \ref{tab:para_all}. We assume $r=3000r_s$ and $q=10$.}
\tablenotetext{a}{The errors represent 1-$\sigma$ uncertainty.}
\tablenotetext{b}{No cooling spectral break is expected in the electron distribution.}
\end{deluxetable*}

\subsection{Fitting Method}
We fit the global X-ray and gamma-ray data simultaneously. However, as the measurement uncertainties of TeV gamma-ray data are large, we do not consider the measurement error for the global fit. Otherwise, the fitting is determined solely by X-ray data whose measurement errors are relatively small. Therefore, we do not determine uncertainties of parameters from the global fits. With those global data, we determine the parameters and the overall spectral shape. Once we find the parameters from the global fits, we refit the model with the {\it Swift}/BAT X-ray spectral data including measurement errors by setting $\eta_g$ as a single free parameter again using the $\chi^2$ minimization technique \citep[e.g.][]{pre92} which allow us to determine uncertainties of $\eta_g$ \footnote{\citet{yan13} have recently developed the Markov Chain Monte Carlo (MCMC) method based on the Bayesian statics for the blazar spectral fit.}. 

The {\it MAXI}/GSC data were not used for the fits. The flux information of the {\it MAXI}/GSC catalog is estimated by assuming a Crab-like spectrum which has a photon index of 2.1 \citep{hir13} which is different from the typical X-ray photon index of HBLs, $\sim3$. The source photon index information is necessary to convert the catalog flux to the true differential flux \citep[see e.g.][]{iso10}. However, the photon index information is not provided in the  {\it MAXI}/GSC catalog. 

For 1ES~0229+200, we do not include the two lowest energy data points of the {\it Fermi}/LAT data which show unusual inverted spectra below $\sim1$~GeV. Gamma-ray induced cascade emission will not make such an inverted spectra at $\sim1$~GeV \citep[e.g.][]{der11,fin15}.  Here, the sky position of 1ES~0229+200 is on the Ecliptic and the moon's path. The Sun and the moon are fairly bright in gamma-ray and the gamma-ray radiation field of the Sun extends several degrees from it \citep{abd11_sun,abd12_moon,ng15}. The flux of 1ES~0229+200 may be contaminated by the solar and/or lunar gamma-ray emission, although those contamination are removed in the 3FGL catalog analysis procedure. Although we tried fitting these two {\it Fermi}/LAT lowest energy data by the synchrotron emission component, we do not find any parameters reproducing these data due to the spectral upper limit at $\sim100$~MeV.

\section{Model parameters of TeV HBLs}
\label{sec:para}

\autoref{fig:SED_v1} and \autoref{fig:SED_v2} show the spectral fitting results of our 13 HBL samples. We show the 37-month averaged {\it MAXI}/GSC, the 70-month averaged {\it Swift}/BAT, the 48-month averaged {\it Fermi}/LAT, and quiescent state IACTs data together with the archival NASA/IPAC Extragalactic Database (NED) data as a reference. We show the integrated flux for the {\it MAXI}/GSC data in these plots, since the photon index information is not available. 

The best-fit SSC models are shown by solid curves. We also show the models having $\gamma_{\rm min}'=1$ and $\eta_g=1$ by dot-dashed and dashed curves, respectively, fixing the other parameters the same as the best-fit model. The fitting results are summarized in Table \ref{tab:para}. The other parameters which are not free parameters but are derived in our self-consistent way are summarized in Table \ref{tab:para_all}. 

Non-relativistic shock acceleration sites such as supernova remnants are known to be in the Bohm limit \citep[e.g.][]{uch07}. However, if we set $\eta_g=1$ (i.e. Bohm limit), the SSC models overproduce the observed hard X-ray fluxes (see \autoref{fig:SED_v1} and \autoref{fig:SED_v2}). Our spectral fits of TeV HBLs give $<\log\eta_g>=4.5\pm0.60$ (Table \ref{tab:para}). This is consistent with past studies of individual objects \citep{ino96,sat08,fin08}. Galactic microquasar jets are also expected to have high $\eta_g$ values as $\sim10^6$ indicated from multi-wavelength spectral fits \citep{tan16}. This high $\eta_g$ value indicates low particle acceleration efficiency implying low magnetic-field-turbulence amplitude. From the VLBI observations, the magnetic field lines are known to be near perpendicular to the shock front \citep[e.g.][]{mar08}. Turbulence are not efficiently generated in quasi-perpendicular shocks and, as a result, such shocks are inefficient to accelerate particles in magnetized relativistic shocks \citep[see e.g.][]{sir13,sir15_review}. Therefore, our result on $\eta_g$ is also consistent with these observational and numerical experiments. As seen in \autoref{fig:SED_v1} and \autoref{fig:SED_v2}, future MeV gamma-ray measurements with the sensitivity limit of $\sim10^{-12}$ erg~cm$^{-2}$~s$^{-1}$ at the MeV gamma-ray bands will determine $\eta_g$ more precisely. With that sensitivity limit, we can expect a few hundred AGN detections \citep{ino15_mev}.

\begin{figure}
 \begin{center}
  \includegraphics[width=9.0cm]{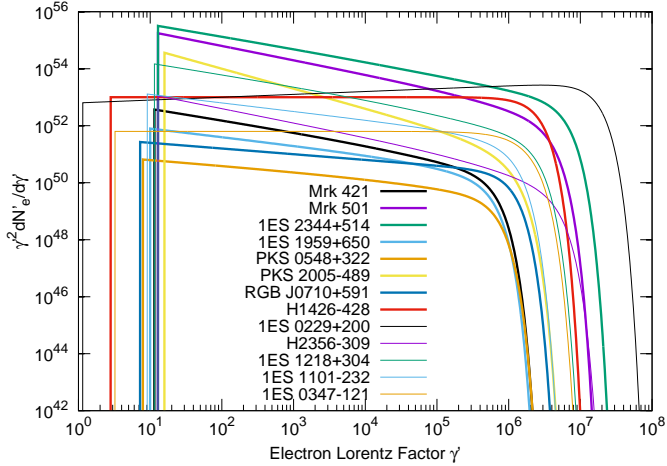} 
 \end{center}
\caption{Electron distribution for our HBL samples in $\gamma'^2dN'_e/d\gamma'(\gamma')$ in the comoving frame. Each curve corresponds to each object as indicated in the figure.}\label{fig:Ne}
\end{figure}

\subsection{Electron Energy Distribution}

\begin{figure}
 \begin{center}
  \includegraphics[width=9.0cm]{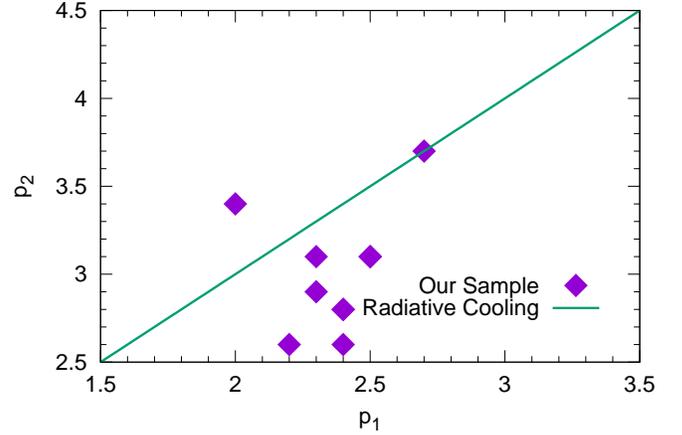} 
 \end{center}
\caption{Relation between the indices of electron distributions. The diagonal line corresponds to the expected relation assuming homogeneous radiative cooling. We do not show the sources in which no cooling spectral break is expected in the electron distribution. }\label{fig:index}
\end{figure}

\autoref{fig:Ne} and \autoref{fig:index} represent the electron spectrum distribution $\gamma'^2dN'_e/d\gamma'$ and the scatter plot of $p_1$ and $p_2$, respectively. In \autoref{fig:index}, we also show the expected $\Delta p$ from the radiative cooling effect \citep[e.g.][]{lon94}.

Our TeV HBL samples have $<p_1>=2.3\pm0.22$ which is in agreement with relativistic shock acceleration theory. In the non-relativistic shock regime, $p_1$ is expected to be 2 \citep{kry77,axf78,bel78,bla78}, whereas it is expected to be $\sim2.2$ in the relativistic shock case \citep{kir00,ach01,kes05,sir15_review}. However, the electromagnetic waves scattering particles is assumed to move with the bulk fluid velocity in those studies. If the scattering wave is slower than the incoming flow, the electron spectrum is expected to become softer. In our samples, PKS~2005-489 has the softest index of $p_1=2.7$. Large-angle scattering in strong turbulent fields would allow the index harder as $p_1\sim1$ \citep[e.g.][]{ste07,aoi08,sum12}, although our results indicate $\eta_g\sim10^{4.5}$ implying weak turbulence.

Among HBLs, it is known that there is a population called as extreme blazars \citep[see e.g.][]{cos01,tan14}. As extreme blazars have hard GeV gamma-ray spectra, it requires hard electron spectra which is naturally expected from stochastic acceleration models \citep[e.g.][]{lef11}. In our samples, RGB~J0710+591, H~1426-428, 1ES~0229+200, 1ES~1101-232, and 1ES~0347-121 are categorized in extreme blazars \citep{tan14}. For those sources, extremely hard electron spectral indies ($p_1\sim1.5$) are not required. Thus, extreme blazars can be explained by in the frame work of the DSA theory. For FSRQs, a hard gamma-ray spectrum has been also reported during a bright gamma-ray flare \citep{hay15}. Although \citet{asa15} recently explained this hard GeV spectrum by the stochastic acceleration model, the spectrum can be explained also in the frame work of the DSA theory considering the fast cooling regime electrons \citep{yan15}. We note that all of our HBL samples are in the slow cooling regime.

Among the sources which require a cooling break in the electron distribution, the average $\Delta p$ is $0.64\pm0.32$. This index relation depends on the inhomogeneity of the acceleration region such as flow geometry, magnetic field strength distribution, matter density, and flow velocity \citep{rey09}. Thus, the departure from the expected relation of $\Delta p=1$ of individual objets may reflect the inhomogeneity of the emitting region in each object. Inhomogeneity of jets is expected considering magnetohydrodynamical instabilities of jets \citep[e.g.][]{miz09,mat13}. X-ray observations also revealed stratified jet structures at kpc-scale \citep{kat06}.

\floattable
\begin{deluxetable*}{lcccccccccccc}
\rotate
 \tablecaption{Parameters derived from spectral fittings\label{tab:para_all}}
\tablehead{
  \colhead{Source} & \colhead{$R'$~[cm]\tablenotemark{a}} & \colhead{$\gamma'_{{\rm min}}$} & \colhead{$\gamma'_{b}$} & \colhead{$\gamma'_{c}$} & \colhead{$P_B$~[erg/s]} & \colhead{$P_e$~[erg/s]} & \colhead{$P_p$~[erg/s]} & \colhead{$P_{\rm rad}$~[erg/s]}& \colhead{$P_{\rm jet}$~[erg/s]} & \colhead{$L_{\rm Edd}$~[erg/s]} & \colhead{$E_{p, \rm max}$~[eV]} & \colhead{$E_{p, \rm max}(\eta_g=1)$~[eV]\tablenotemark{c}}\
 }
\startdata
Mrk~421 	&	5.7$\times10^{15}$	&	11	&	1.6$\times10^{5}$	&	5.0$\times10^{5}$	&	4.5$\times10^{42}$	&	4.5$\times10^{44}$	&	4.2$\times10^{45}$	&	2.8$\times10^{42}$	&	4.6$\times10^{45}$	&	2.5$\times10^{46}$	&	6.4$\times10^{13}$	&	7.3$\times10^{18}$\\
Mrk~501 	&	8.7$\times10^{16}$	&	13	&	1.5$\times10^{6}$	&	3.1$\times10^{6}$	&	1.7$\times10^{42}$	&	4.1$\times10^{45}$	&	4.0$\times10^{46}$	&	3.7$\times10^{42}$	&	4.5$\times10^{46}$	&	2.0$\times10^{47}$	&	1.1$\times10^{14}$	&	4.6$\times10^{18}$\\
1ES~2344+514 	&	9.6$\times10^{16}$	&	13	&	3.0$\times10^{5}$	&	4.8$\times10^{6}$	&	7.4$\times10^{41}$	&	9.5$\times10^{44}$	&	8.5$\times10^{45}$	&	9.0$\times10^{42}$	&	9.4$\times10^{45}$	&	7.9$\times10^{46}$	&	3.0$\times10^{14}$	&	3.0$\times10^{18}$\\
1ES~1959+650 	&	3.6$\times10^{15}$	&	10	&	2.6$\times10^{5}$	&	4.7$\times10^{5}$	&	1.8$\times10^{42}$	&	2.3$\times10^{44}$	&	2.1$\times10^{45}$	&	1.5$\times10^{42}$	&	2.3$\times10^{45}$	&	1.6$\times10^{46}$	&	3.5$\times10^{13}$	&	4.6$\times10^{18}$\\
PKS~0548-322 	&	3.1$\times10^{15}$	&	7.9	&	3.0$\times10^{5}$	&	5.3$\times10^{5}$	&	2.3$\times10^{42}$	&	5.3$\times10^{43}$	&	5.0$\times10^{44}$	&	4.6$\times10^{41}$	&	5.5$\times10^{44}$	&	1.8$\times10^{46}$	&	5.1$\times10^{13}$	&	5.3$\times10^{18}$\\
PKS~2005-489 	&	3.2$\times10^{16}$	&	16	&	6.6$\times10^{5}$	&	1.1$\times10^{6}$	&	6.0$\times10^{42}$	&	5.5$\times10^{45}$	&	5.3$\times10^{46}$	&	1.9$\times10^{42}$	&	5.9$\times10^{46}$	&	1.4$\times10^{47}$	&	7.0$\times10^{13}$	&	8.5$\times10^{18}$\\
RGB~J0710+591 	&	3.4$\times10^{15}$	&	7.1	&	-\tablenotemark{b}	&	8.6$\times10^{5}$	&	4.9$\times10^{41}$	&	2.9$\times10^{44}$	&	2.8$\times10^{45}$	&	1.1$\times10^{42}$	&	3.1$\times10^{45}$	&	2.3$\times10^{46}$	&	7.0$\times10^{13}$	&	2.4$\times10^{18}$\\
H~1426+428 	&	5.4$\times10^{16}$	&	2.8	&	-\tablenotemark{b}	&	2.0$\times10^{6}$	&	1.2$\times10^{42}$	&	4.6$\times10^{44}$	&	4.5$\times10^{45}$	&	7.4$\times10^{42}$	&	5.0$\times10^{45}$	&	1.7$\times10^{47}$	&	7.5$\times10^{13}$	&	3.8$\times10^{18}$\\
1ES~0229+200 	&	5.9$\times10^{16}$	&	1.1	&	-\tablenotemark{b}  	&	1.3$\times10^{7}$	&	3.2$\times10^{40}$	&	1.1$\times10^{45}$	&	9.7$\times10^{45}$	&	3.8$\times10^{42}$	&	1.1$\times10^{46}$	&	2.2$\times10^{47}$	&	5.7$\times10^{14}$	&	6.2$\times10^{17}$\\
H~2356-309 	&	8.7$\times10^{15}$	&	13	&	8.5$\times10^{5}$	&	3.8$\times10^{6}$	&	2.3$\times10^{42}$	&	1.6$\times10^{45}$	&	1.5$\times10^{46}$	&	2.3$\times10^{42}$	&	1.7$\times10^{46}$	&	5.0$\times10^{46}$	&	9.6$\times10^{14}$	&	5.3$\times10^{18}$\\
1ES~1218+304 	&	1.8$\times10^{16}$	&	11	&	2.1$\times10^{5}$	&	1.9$\times10^{6}$	&	2.1$\times10^{42}$	&	2.4$\times10^{45}$	&	2.2$\times10^{46}$	&	2.3$\times10^{43}$	&	2.4$\times10^{46}$	&	4.8$\times10^{46}$	&	2.2$\times10^{14}$	&	5.1$\times10^{18}$\\
1ES~1101-232 	&	2.2$\times10^{16}$	&	8.9	&	9.5$\times10^{5}$	&	1.0$\times10^{6}$	&	5.2$\times10^{42}$	&	1.2$\times10^{45}$	&	1.2$\times10^{46}$	&	3.4$\times10^{42}$	&	1.3$\times10^{46}$	&	1.3$\times10^{47}$	&	6.9$\times10^{13}$	&	7.9$\times10^{18}$\\
1ES~0347-121 	&	1.3$\times10^{16}$	&	3.2	&	1.6$\times10^{6}$	&	1.8$\times10^{6}$	&	1.1$\times10^{42}$	&	2.0$\times10^{44}$	&	1.9$\times10^{45}$	&	4.1$\times10^{42}$	&	2.1$\times10^{45}$	&	5.6$\times10^{46}$	&	9.0$\times10^{13}$	&	3.5$\times10^{18}$\\
\enddata
\tablecomments{Parameters derived from the best-fit parameters (Table \ref{tab:para}). Quantities in the jet comoving frame of the emitting region are primed. The other quantities are in the stationary frame. See the text for the detail how the parameters are derived.}
\tablenotetext{a}{We assume $r=3000r_s$ and $q=10$.}
\tablenotetext{b}{As $\gamma'_{b}>\gamma'_{\rm c}$, we expect no cooling spectral break in the electron spectrum.}
\tablenotetext{c}{We set $\eta_g=1$ but keep the other best-fit parameters unchanged.}
\end{deluxetable*}

\begin{figure}
 \begin{center}
  \includegraphics[width=9.0cm]{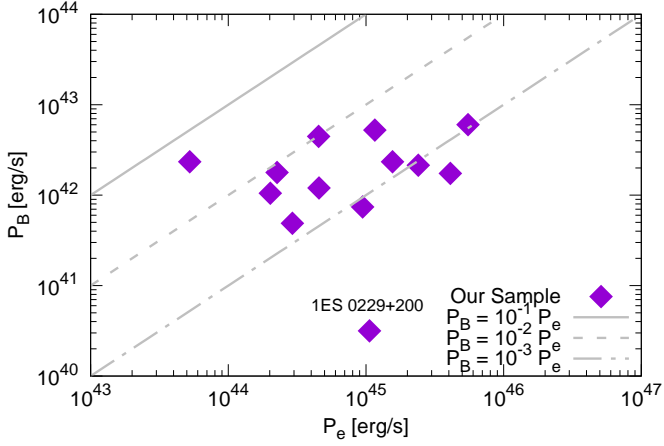} 
 \end{center}
\caption{Magnetic field power $P_B$ as a function of electron power $P_e$ for our TeV HBL samples. Diamonds are for our samples. Solid, dashed, and dot-dashed line corresponds to the $P_B/P_e = 0.1$, $P_B/P_e = 0.01$, and $P_B/P_e=0.001$, respectively. 1ES~0229+200 showing extremely low $P_B$ is indicated in the plot. We assume $r=3000r_s$ and $q=10$.}\label{fig:Pe_PB}
\end{figure}

\section{Jet Power and Jet Baryon Mass Loading Efficiency}
\label{sec:power}

Power of relativistic jets is a powerful tool for the understanding of the jet physics. Energetics of blazar jets has been studied in literature \citep[e.g.][]{tav98,cel08,tav10,ghi10,zha12,ghi14,tav15}. However, hard X-ray data was lacking which is essential to determine the highest energy of electrons in the HBL spectral fitting. Furthermore, our method evaluates $\gamma'_{\rm min}$, $\gamma'_{b}$, and $\gamma'_{c}$ from the other parameters listed in Table \ref{tab:para} self-consistently in the DSA theory. In this section, we evaluate the power of relativistic jets from our spectral fits.

\begin{figure}
 \begin{center}
  \includegraphics[width=9.0cm]{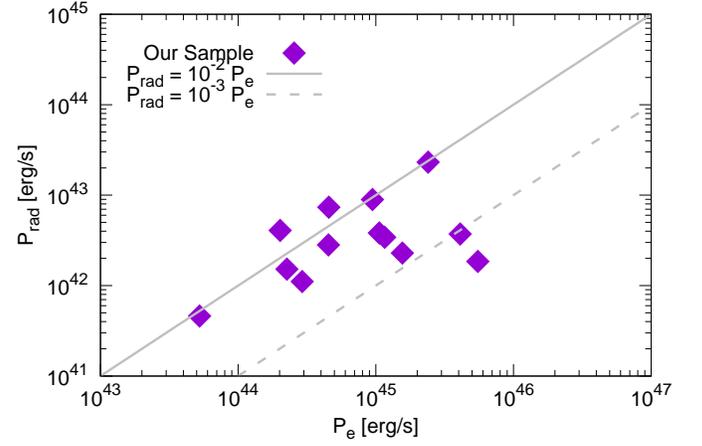} 
 \end{center}
\caption{Radiation power $P_{\rm rad}$ as a function of electron power $P_e$ for our TeV HBL samples. Diamonds are for our samples. Solid and dashed line corresponds to the $P_{\rm rad}/P_e = 0.01$ and 0.001, respectively. We assume $r=3000r_s$ and $q=10$.}\label{fig:Pe_Prad}
\end{figure}

The comoving magnetic field energy in the blob is given by
\begin{equation}
W_B' = V'U_B = \frac{R'^3B^2}{6},
\end{equation}
where $V'$ is the volume of the emitting blob $4\pi R'^3/3$. 

The total comoving electrons and positrons energy is given by
\begin{equation}
W_e' = m_ec^2 \int_{\gamma_{\rm min}'}^{\infty} d\gamma' \gamma' N_e'(\gamma).
\end{equation}

The power of each component in the stationary frame is given by
\begin{equation}
P = 2\pi R'^2 \Gamma^2 c W'/V' = \frac{3\delta^2cW'}{2R'},
\end{equation}
where we omit $\beta$ and the factor of 2 in the second term is due to the twin jet. 

\autoref{fig:Pe_PB} shows the ratio between $P_B$ and $P_e$. The ratio is the same as in energy density. The average ratio is $<P_B/P_e> = 6.2\times10^{-3}\pm1.1\times10^{-2}$. Such low magnetic field energy comparing to electrons makes reconnection acceleration inefficient \citep{sir15}. 1ES~0229+200 requires extremely low magnetic field strength as $P_B/P_e=3.0\times10^{-6}$ comparing to the others. Departure from equipartition is consistent with previous studies \citep[e.g.][]{tav98,cel08,tav10,ghi10,zha12,tav15}. On the contrary, FSRQs are known to be in near equipartition $P_B\simeq P_e$ \citep{cel08,ghi10,der15}. 

 \citet{tav15} have recently proposed that structured jets (a fast spine surrounded by a slow sheath) may ease the equipartition problem of BL~Lacs because of radiative interaction between two layers. However, it is beyond the scope to fit the data with the structured jet model \citep{ghi05} in this paper.

\begin{figure}
 \begin{center}
  \includegraphics[width=9.0cm]{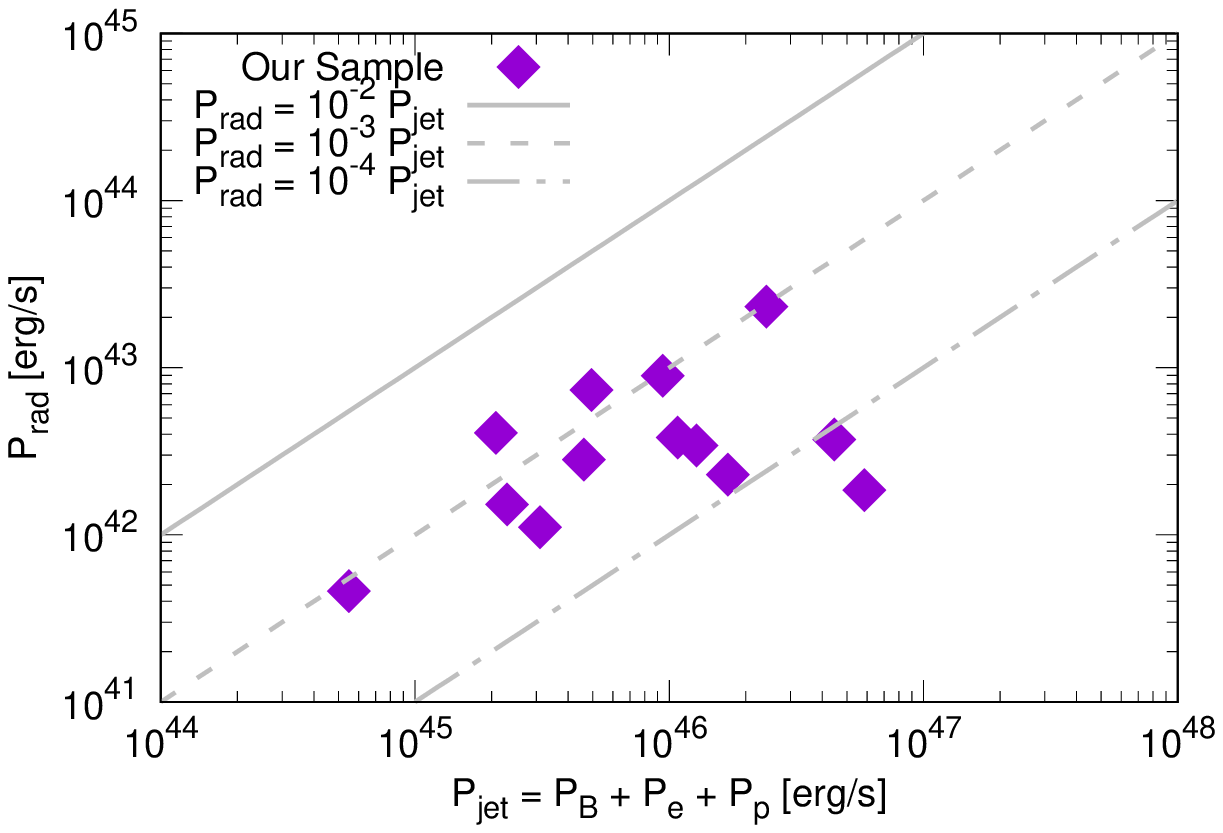} 
 \end{center}
\caption{Radiation power $P_{\rm rad}$ as a function of total jet power $P_{\rm jet}=P_B+P_e+P_p$ for our TeV HBL samples. Diamonds are for our samples. Solid, dashed, and dot-dashed line corresponds to the $P_{\rm rad}/P_{\rm jet} = 10^{-2}$, $10^{-3}$, and $10^{-4}$, respectively. We assume $r=3000r_s$ and $q=10$.}\label{fig:Prad_Pjet}
\end{figure}

The total radiating power is given by
\begin{equation}
P_{\rm rad} = \frac{4\pi d_L(z)^2}{2\Gamma^2} \int_0^\infty d\nu f_\nu/\nu,
\end{equation}
where $d_L$ is the luminosity distance to the source at a redshift $z$, $1/2\Gamma^2$ is due to the twin jet, $f_\nu$ is the observed flux in the unit of [${\rm erg/cm^2/s}$], and $\nu$ is the photon frequency.  We present the comparison between $P_e$ and $P_{\rm rad}$ of our TeV HBL samples in \autoref{fig:Pe_Prad}. The average ratio is $<P_{\rm rad}/P_{\rm e}>=(6.9\pm5.7)\times10^{-3}$. 

The total jet power is given by
\begin{equation}
P_{\rm jet} = P_B + P_e + P_p,
\end{equation}
where $P_p$ is the proton power. We ignore the photon power which is negligible (Figures \ref{fig:Pe_Prad} and \ref{fig:Prad_Pjet}). Although protons are not responsible for the photon emission in the leptonic SSC model, we are able to estimate the proton energy as (see Equation \ref{eq:gmin})
\begin{equation}
W_p' \simeq \frac{W_e'}{\epsilon_e}=\frac{\Gamma_{\rm sh} m_pc^2 \int_{\gamma_{\rm min}'}^{\infty} d\gamma' N_e'(\gamma)}{q},
\end{equation}
where $m_p$ is the proton rest mass. As in Equation \ref{eq:gmin}, we assume one proton per $q$ radiating leptons. Since we define leptons carry 10\% of the shocked energy, the main energy carrier of the jet is protons by definition in this paper. In Section \ref{subsec:diff_gmin}, we consider the other case in which we assume  cold protons only and $\gamma'_{\rm min}$ determined by the mass ratio between a proton and an electron.

\autoref{fig:Prad_Pjet} shows the ratio between $P_{\rm rad}$ and $P_{\rm jet}$ of TeV HBLs. The average ratio is $<P_{\rm rad}/P_{\rm jet}>=(6.7\pm5.5)\times10^{-4}$. This ratio represents the radiative efficiency of HBL jets $\epsilon_{\rm rad,jet}$. $\epsilon_{\rm rad,jet}\sim0.1$ is known for luminous blazars and gamma-ray bursts \citep{nem12,ghi14}. However, it is assumed that protons are cold in these papers. Therefore, for a fair comparison, we need to evaluate the proton power of those different object classes in the same way. We will show the case for the cold proton fit of our samples in Section \ref{subsec:diff_gmin}.

\subsection{Relation between Accretion Inflows and Jet Outflows}

\begin{figure}
 \begin{center}
  \includegraphics[width=9.0cm]{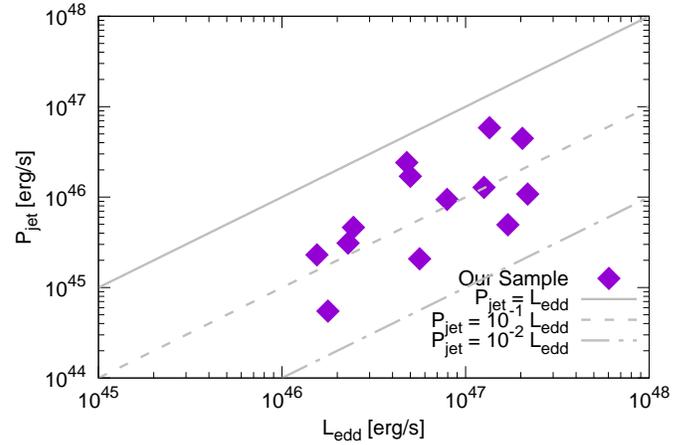} 
 \end{center}
\caption{Jet power $P_{\rm jet}$ as a function of the Eddington luminosities $L_{\rm Edd}$ for our TeV HBL samples. Diamonds are for our samples. Solid, dashed, and dot-dashed line corresponds to $P_{\rm jet}/L_{\rm Edd} = 1$, $10^{-1}$, and 10$^{-2}$, respectively. We assume $r=3000r_s$ and $q=10$.}\label{fig:L_edd_Pjet}
\end{figure}

\autoref{fig:L_edd_Pjet} shows the ratio between $P_{\rm jet}$ and $L_{\rm Edd}$, $L_{\rm Edd} \simeq 1.3\times10^{46} (M_{\rm BH}/10^8M_\odot)$ erg s$^{-1}$. The average ratio is $<P_{\rm jet}/L_{\rm Edd}>=0.18\pm0.15$. This tight relation suggests the correlation between the jet power and the SMBH mass for HBLs. 

Accretion rate in BL Lacs is known to be as low as in the RIAF regime \citep[e.g.][]{ghi10} which is sometimes called as advection dominated accretion flow \citep[ADAF;][]{kat98,kat08}. The mass accretion rate of HBLs is typically about an order of $\dot{m}_{\rm acc}\simeq1.2\times10^{-2}$ of the Eddington mass accretion rate  \citep{wan02}. The accretion rate is estimated from the disk luminosities using the self-similar solution for ADAF disks \citep{mah97} where the inner disk region dominates the radiation. In \citet{wan02}, the accretion efficiency in converting matter to energy was set to be $\epsilon_{\rm rad,acc}=0.1$ \citep{fra92} and disk luminosities of HBLs were evaluated by using line luminosities. The disk mass accretion rate is given by $\dot{M}_{\rm acc}=\dot{m}_{\rm acc}L_{\rm Edd}/\epsilon_{\rm rad,acc} c^2$. Therefore, we have the jet production efficiency as
\begin{eqnarray}
\eta_{\rm jet}&\equiv&\frac{P_{\rm jet}}{\dot{M}_{\rm acc}c^2}=\frac{\epsilon_{\rm rad,acc}}{\dot{m}_{\rm acc}}\frac{P_{\rm jet}}{L_{\rm Edd}}\\
&\simeq&1.5\left(\frac{\epsilon_{\rm rad,acc}}{0.1}\right) \left(\frac{\dot{m}_{\rm acc}}{1.2\times10^{-2}}\right)^{-1} \left(\frac{P_{\rm jet}/L_{\rm Edd}}{0.18}\right).
\end{eqnarray}
$\eta_{\rm jet}>1$ implies a part of the SMBH energy would be extracted to launch a relativistic jet. Luminous blazars also have $\eta_{\rm jet}\sim 1.4$ \citep{ghi14}, although $q=1$ is assumed.

Following the recent numerical studies \citep{tch11,mck12}, powerful relativistic jets are launched in the magnetically arrested/choked accretion flows. Extracted jet power by the rotation of BHs threaded by magnetic fields, so-called the BZ power, is given by \citep{bla77,tch10,tch11}
\begin{eqnarray}
P_{\rm BZ} &=& 4.0\times10^{-3}\frac{1}{c}\Omega_{\rm H}^2 \Phi_{\rm BH}f(\Omega_{\rm H})\\
&\simeq& 10 \left(\frac{\phi_{\rm BH}}{50}\right)^2x_a^2f(x_a)\dot{M}_{\rm acc}c^2,
\end{eqnarray}
where $\Omega_{\rm H}=ac/2r_{\rm H}$ is the angular frequency of the BH horizon, $\Phi_{\rm BH}$ is the net magnetic field flux accumulated in the central region, $x_a\equiv r_g\Omega_{\rm H}/c$, and $f(x_a)\approx1+1.38x_a^2-9.2x_a^4$. $a\equiv J_{\rm BH}/J_{\rm BH, max}=cJ_{\rm BH}/GM_{\rm BH}^2$ is the dimensionless BH spin parameter, $r_{\rm H}=r_g(1+\sqrt{1-a^2})$ is the horizon radius, $r_g = GM_{\rm BH}/c^2$ is the gravitational radius of the BH. $\phi_{\rm BH}=\Phi_{\rm BH}/\sqrt{\dot{M}_{\rm acc}r_g^2c}$ is the dimensionless magnetic flux threading the BH and is typically on the order of 50 \citep{mck12}. This gives $\eta_{\rm jet, BZ}\equiv P_{\rm BZ}/\dot{M}_{\rm acc}c^2 \simeq10 (\phi_{\rm BH}/50)^2x_a^2f(x_a)$. Given $\eta_{\rm jet}\sim1.5$, the spin parameter $a$ would be 0.97 implying a rapidly spinning BH neglecting a contribution from accretion flow onto the jet power.

Here, the mass outflow rate of the jet can be described as $\dot{M}_{\rm jet}\gtrsim P_{\rm jet}/\Gamma c^2$ because the magnetic field power is negligible (\autoref{fig:Pe_PB}) and we can take into account relativistic particle mass only. Therefore, the mass loading efficiency from accretion flows to jet outflows is given as
\begin{eqnarray}
\xi_{\rm jet}&\equiv&\frac{\dot{M}_{\rm jet}}{\dot{M}_{\rm acc}} \gtrsim \frac{\epsilon_{\rm rad,acc} P_{\rm jet}}{\Gamma \dot{m}_{\rm acc} L_{\rm Edd}}\\
 \gtrsim&5\times&10^{-2} \left(\frac{\epsilon_{\rm rad,acc}}{0.1}\right) \left(\frac{\dot{m}_{\rm acc}}{1.2\times10^{-2}}\right)^{-1}\left(\frac{\Gamma}{30}\right)^{-1} \left(\frac{P_{\rm jet}/L_{\rm Edd}}{0.18}\right),
\end{eqnarray}
where we take the average beaming factor of our samples (Table \ref{tab:para}). 

$\gtrsim$5\% of accreted mass are ejected as outflow jet for TeV HBLs. This implies that the most of accreted mass is stored in the central SMBH or ejected as wide-opening angle disk wind outflows, although the derived fraction is still highly uncertain due to various assumptions. As the accretion rate is low, such mass accretion will not significantly contribute to the SMBH mass growth. RIAFs also inevitably generate wide-angle disk outflows (not collimated jet) \citep[e.g.][]{bla99} which eject $\sim30$\% of accreted mass \citep[see e.g.][]{tot06_riaf}. Furthermore, radio-loud galaxies are known to generate ultra-fast outflow whose rate is comparable to the accretion rate \citep{tom10,tom14}.

\begin{figure}
 \begin{center}
  \includegraphics[width=9.0cm]{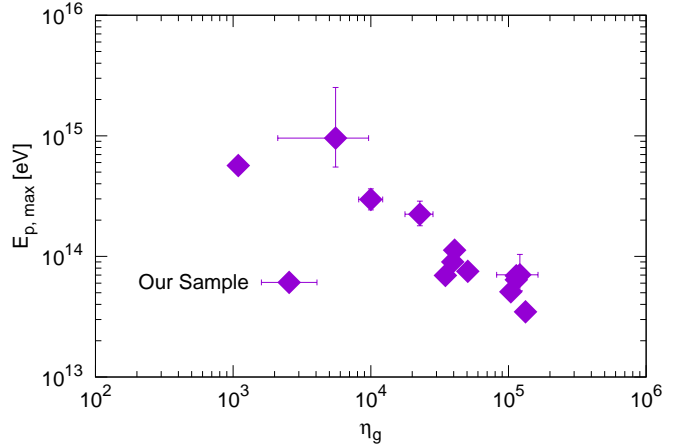} 
 \end{center}
\caption{Maximum possible proton energies of our TeV HBL samples as a function of gyrofactors $\eta_g$. The proton energies are shown in the stationary frame. The error bars represent 1-$\sigma$ uncertainty of $\eta_g$. For some sources, uncertainties are so small that error bars are not apparently seen in the plot.}\label{fig:emax}
\end{figure}

Recent 3D global GRMHD simulations for the jet launching from a RIAF disk indicates $\xi_{\rm jet}\sim0.12$ for a rapidly spinning BH $a=0.94$ and $\xi_{\rm jet}\sim0$ for a slowly spinning BH $a=0.5$ \citep{mck12} where $a$ is the BH spin parameter and $\dot{M}_{\rm jet}$ and $\dot{M}_{\rm acc}$ are evaluated at $50r_s$ from the central SMBH and at the BH horizon. When we take $\dot{M}_{\rm acc}$ at $50r_s$ from the central SMBH, $\xi_{\rm jet}$ will be $\sim0.028$ and $\sim0$ for $a=0.94$ and $a=0.5$, respectively \citep{mck12}. Our results suggest that HBLs are in between $a=0.94$ and $a=0.5$. 

\citet{tom12} analytically estimate $\xi_{\rm jet}\sim6\times10^{-4}$ for $M_{\rm BH}=10^8M_\odot$ considering relativistic neutron injection from the accretion flows to the jet. However, it requires high neutron luminosities from the disk which is inconsistent with the study of disk structure studies of RIAFs \citep{kim14,kim15}. As we have discussed $\eta_{\rm jet}$ and $\xi_{\rm jet}$ using the average parameters of the 13 TeV HBLs and on various assumptions, further detailed  studies of relation between accretion inflows and jet outflows is required for quantitative discussions.

\begin{figure}
 \begin{center}
  \includegraphics[width=9.0cm]{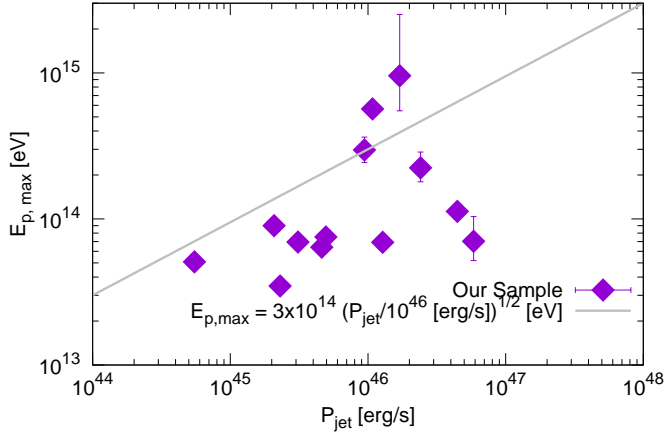} 
 \end{center}
\caption{Maximum possible proton energies of our TeV HBL samples as a function of jet power. The proton energies are shown in the stationary frame. The error bars represent 1-$\sigma$ uncertainty. Solid line represents the relation of $E_{p, {\rm max}} \simeq 3\times10^{14} (f_B/6\times10^{-4})^{1/2} ({P_{\rm jet}}/{10^{46}~{\rm erg/s}})^{1/2} ({\eta_g}/{10^{4.5}})^{-1}~[\rm eV].$}\label{fig:emax_Pjet}
\end{figure}

\section{Particle Acceleration Efficiency and Maximum Proton Energy}
\label{sec:PAE}
Since we obtain the blob size, the magnetic field strength, and the particle acceleration efficiency from the multi-wavelength spectral fits, we are able to evaluate the maximum possible proton energy in the blazar zone following the Hillas argument \citep{hil84}. The maximum proton energy in the stationary frame is given as 
\begin{eqnarray}
E_{p, {\rm max}} &=& \delta eBR'/\eta_g \\ \label{eq:emax}
\simeq 9\times&10^{18}& \left(\frac{\delta}{10}\right) \left(\frac{B}{0.1~{\rm G}}\right) \left(\frac{R'}{3\times10^{16}~{\rm cm}}\right) \left(\frac{\eta_g}{1}\right)^{-1}~[\rm eV].
\end{eqnarray}
If the acceleration is near the Bohm limit, UHECRs can be accelerated in the blazar zone in principle. 

\autoref{fig:emax} shows the maximum proton energies of our TeV HBL samples based on parameters obtained from our fits. As $\eta_g$ is $\sim10^{4.5}$, the maximum possible proton energy is $\sim10^{14-15}~{\rm eV}$. The energy is much lower than the energy of UHECRs even if we consider the uncertainties of $\eta_g$ or iron element. These results suggest that the blazar zone of low power AGN jets could not be the UHECR acceleration sites under the assumption of the DSA scenario. This also implies blazars may not be responsible for the IceCube detected TeV--PeV neutrinos \citep{aar13,aar14}. As discussed in \citet{mur14}, $\eta_g\lesssim10^4$ is required to be efficient neutrino emitters, if we assume the first-order Fermi acceleration process. However, blazars like 1ES~0229+200 which has $\eta_g\sim10^3$ will be able to emit TeV--PeV neutrinos. 

$E_{p,{\rm max}}$ should depend on the jet power \citep[e.g.][]{der10}. \autoref{fig:emax_Pjet} shows the dependence of $E_{p,{\rm max}}$ on the jet power $P_{\rm jet}$. The magnetic field energy density can be written as 
\begin{equation}
U_B=f_B \frac{P_{\rm jet}}{2\pi R'^2 \Gamma^2 c}
\end{equation}
where $f_B$ corresponds to the fraction of the magnetic field energy in the jet. Our analysis indicates $f_B\simeq 6\times10^{-4}$. Thus, from Equation \ref{eq:emax}, we have
\begin{eqnarray}
E_{p, {\rm max}} &=& \frac{2e}{\eta_g }\sqrt{\left(\frac{f_BP_{\rm jet}}{c} \right)}\\
\simeq 3\times&10^{14}& \left(\frac{f_B}{6\times10^{-4}}\right)^{1/2} \left(\frac{P_{\rm jet}}{10^{46}~{\rm erg/s}}\right)^{1/2} \left(\frac{\eta_g}{10^{4.5}}\right)^{-1}~[\rm eV].
\end{eqnarray}
This expected dependence is also shown in \autoref{fig:emax_Pjet}.

We assumed that protons and electrons have the same acceleration efficiencies. However, the gyro radius of the protons is $\sim m_p/m_e$ larger than electrons. This means that protons are accelerated by different part of the wave turbulence spectrum from electrons. Thus, large scale turbulences such as those generated by magnetohydrodynamical instabilities during jet propagation \citep[e.g.][]{miz09,guo12,mat13} would reaccelerate particles to higher energies. In Table \ref{tab:para_all}, we show $E_{p,{\rm max}}$ setting $\eta_g=1$ which gives the most optimistic maximum proton energy. With $\eta_g=1$, $<E_{p,{\rm max}}>\sim (2.1\pm2.6)\times10^{18}$~eV. To be UHECR accelerators, in this case, the iron composition will be necessary. However, the composition of UHECRs is still under discussion \citep[e.g.][]{abr10_comp,abb10,abb15}

\begin{figure}
 \begin{center}
  \includegraphics[width=9.0cm]{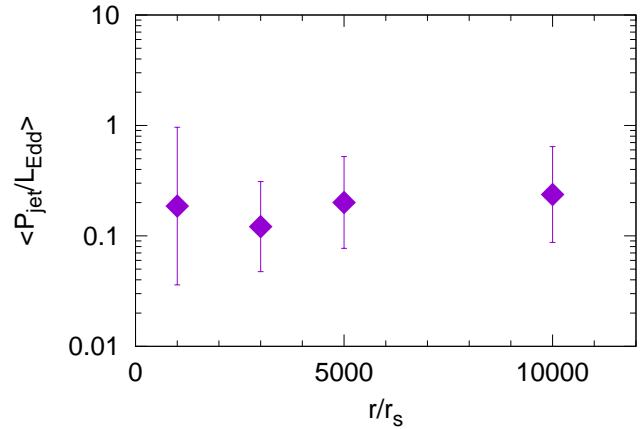} 
 \end{center}
\caption{Dependence of $<P_{\rm jet}/L_{\rm Edd}>$ on the location of the emitting region of our 13 TeV HBL samples. The error bars represent 1-$\sigma$ uncertainty.}\label{fig:rsch_xi}
\end{figure}

The second-order Fermi acceleration scenarios can also reproduce various blazar SEDs \citep[e.g.][]{yan13,asa14}. The second-order Fermi acceleration process will require the lower $\eta_{\rm acc}$ than that required for the first-order process \citep[see Figures 6 and 7 of][]{der14_dmi}. Thus, higher maximum proton energy and higher neutrino energy would be expected for the second-order process.

\section{Discussion}
\label{sec:dis}

\subsection{Dependence on the Location of the Emitting Region}
\label{subsec:diff_r}

\begin{figure}
 \begin{center}
  \includegraphics[width=9.0cm]{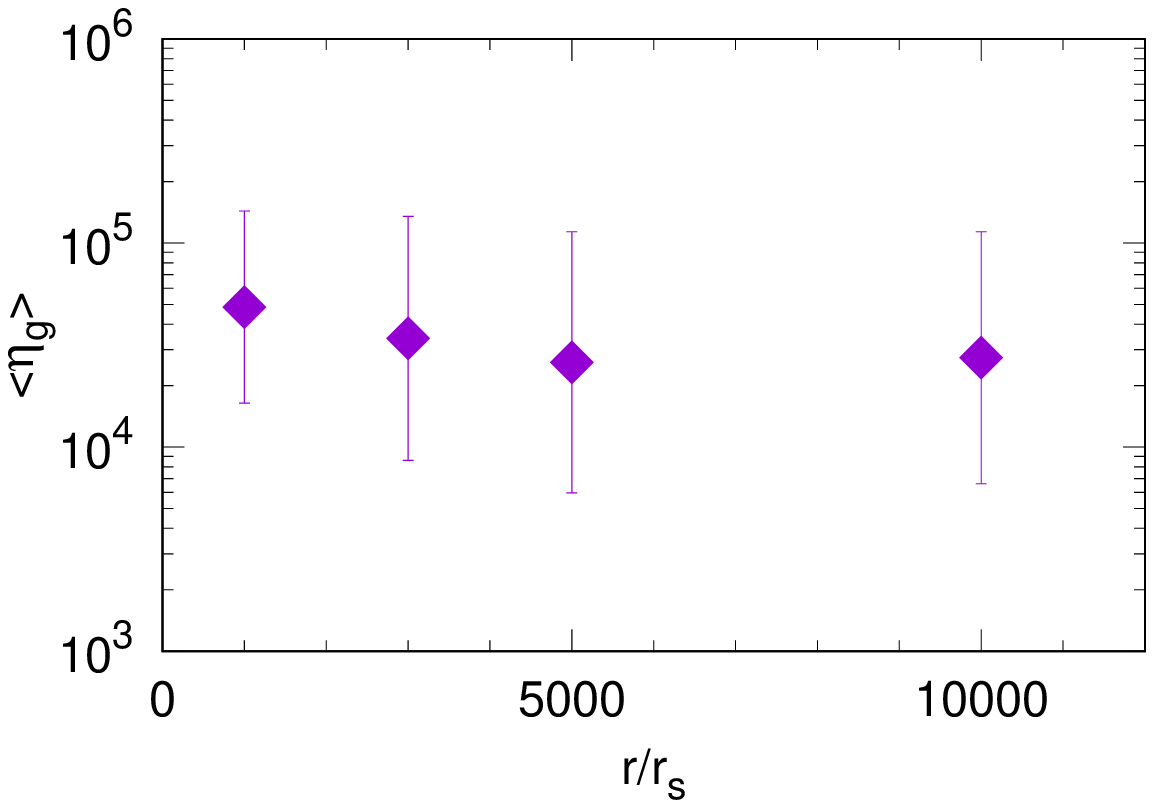} 
 \end{center}
\caption{Same as \autoref{fig:rsch_xi} but for $<\eta_g>$.}\label{fig:rsch_eta}
\end{figure}

We assume the location of the emitting region is at $r=3\times10^3r_s$ from the central SMBH. However, the location of emission sites in blazar jets is a long standing issue in the astrophysics. We have further tested the cases for different locations $r=10^3$, $5\times10^3$, and $10^4$ $r_s$. \autoref{fig:rsch_xi} and \autoref{fig:rsch_eta} shows the dependence of $<P_{\rm jet}/L_{\rm Edd}>$ and $<\eta_g>$ on the location of the emitting region, respectively. The other assumptions are unchanged. The uncertainties in the plots represents the standard deviation of each parameter for each location. We do not see clear differences in inferred values in both plots by the choice of the location of the emitting region. Therefore, our conclusions will not be severely affected by our assumption of $r=3\times10^3r_s$. We note that FSRQs may have emitting regions further away \citep[see e.g.][for the case of 4C~+21.35]{tan11}.

\subsection{Dependence on the Pair Content}
\label{subsec:diff_q}
In this paper, we assume 10 leptons per one proton in the jet ($q=10$). The jet energetics arguments strongly depend on the value of $q$. However, as discussed above (see Section \ref{sec:model}), the jet composition is not well determined \citep[see e.g.][]{kat08_1510}. We further test the different values of $q=1, 20$, and 50. $q=1$ represents the one electron per one proton jet model. 

\begin{figure}
 \begin{center}
  \includegraphics[width=9.0cm]{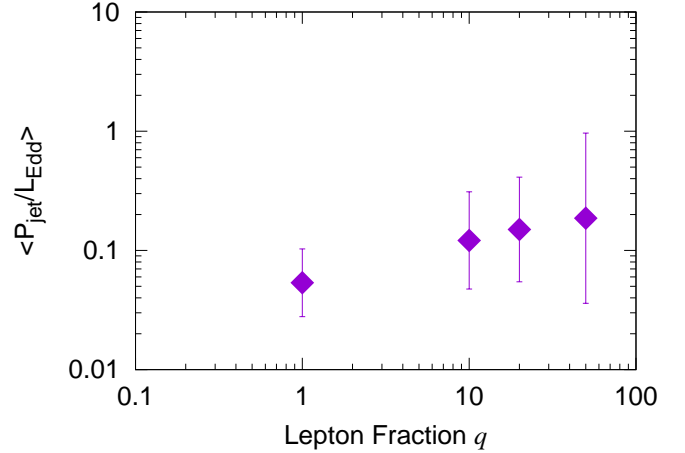} 
 \end{center}
\caption{Dependence of $<P_{\rm jet}/L_{\rm Edd}>$ on the lepton fraction against proton of our 13 TeV HBL samples. The error bars represent statistical 1-$\sigma$ uncertainty.}\label{fig:q_Pjet}
\end{figure}

\autoref{fig:q_Pjet} shows the dependence of $<P_{\rm jet}/L_{\rm Edd}>$ on the lepton fraction. The other assumptions are unchanged. The uncertainties in the plots represents the standard deviation of each parameter. $<P_{\rm jet}/L_{\rm Edd}>$ gradually increase with $q$, although the scatter is large. This is understood as follows. The minimum electron Lorentz factor is determined by the total energy ratio of leptons against protons (\autoref{eq:gmin}). If $p_1>2$, $\gamma'_{\rm min}\simeq370(p_1-2)/q(p_1-1)$ with $\epsilon_e=0.1$ and $\Gamma_{\rm sh}=2$. Since $<p_1>\sim2.3$, $\gamma'_{\rm min}\sim85/q$. Therefore, as $q$ increases, $\gamma'_{\rm min}$ decreases. This results in the increase of the total number of leptons and protons. Therefore, jet power increases with $q$ in our modelling.

\subsection{Different Minimum Electron Lorentz Factor}
\label{subsec:diff_gmin}

To determine the minimum Lorentz factor $\gamma'_{\rm min}$, we assume that 10\% of the shocked energy goes into lepton acceleration (Equation \ref{eq:gmin}). Here, the shock thickness is expected to be of the order of the gyroradius of protons. Then, to cross the shock,
the required minimum electron Lorentz factors would be approximately determined by the mass ratio between a proton and an electron as
\begin{equation}
\label{eq:gmin_ratio}
\gamma'_{\rm min}\sim \frac{m_p}{m_e},
\end{equation}
although we need to take into account various microphysics in the shock to determine the true $\gamma'_{\rm min}$. For the terminal shocks of quasar jets, $\gamma'_{\rm min}$ is known to be $\sim m_p/m_e$ \citep{sta07}.

In this case, the proton energy in the jet can not be constrained from leptonic SSC spectral modelling. Assuming cold protons only, the proton energy can be evaluated as
\begin{equation}
W_p' = \frac{m_pc^2 \int_{\gamma_{\rm min}'}^{\infty} d\gamma' N_e'(\gamma_e')}{q}.
\end{equation} 
This power estimation gives the minimum proton energy in the emitting plasma. Such cold protons have been assumed in the past studies but setting $\gamma'_{\rm min}$ as a free parameter \citep[e.g.][]{ghi14}. 

\begin{figure}
 \begin{center}
  \includegraphics[width=9.0cm]{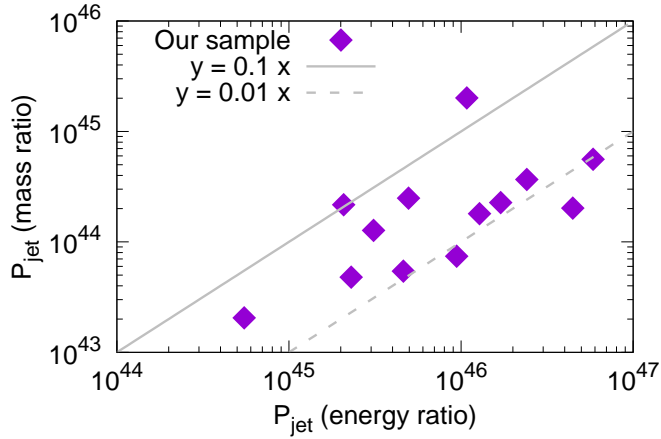} 
 \end{center}
\caption{Comparison of $P_{\rm jet}$ of our 13 HBL samples between the models in which $\gamma'_{\rm min}$ is determined by the shocked energy  (Equation \ref{eq:gmin}) or the mass ratio of $m_p/m_e$ (Equation \ref{eq:gmin_ratio}).Solid and dashed line corresponds to the ($P_{\rm jet}$ [mass ratio]) /  ($P_{\rm jet}$ [energy ratio]) = 0.1 and 0.01, respectively. }\label{fig:gmin_Pjet}
\end{figure}

To investigate how our results are affected by the assumption on $\gamma'_{\rm min}$. We fit our HBL samples again but with $\gamma'_{\rm min}$ defined by the mass ratio. The other assumptions remain unchanged. \autoref{fig:gmin_Pjet} shows the comparison of $P_{\rm jet}$ based on models in which $\gamma'_{\rm min}$ is determined by the shocked energy  (Equation \ref{eq:gmin}) or the mass ratio of $m_p/m_e$ (Equation \ref{eq:gmin_ratio}). In our fiducial model, we have $<\gamma'_{\rm min}>=9.0\pm4.2$ (Table \ref{tab:para_all}) which is 200 times smaller than $m_p/m_e\sim1.8\times10^3$. Thus, $P_{\rm jet}$ estimated in the fiducial model is expected to be higher than that estimated with $\gamma'_{\rm min}=m_p/m_e$.  

By assuming $\gamma'_{\rm min}=m_p/m_e$, we have $<\epsilon_{\rm rad,jet}>=(4.5\pm7.3)\times10^{-2}$ for $q=10$ and $(3.1\pm5.0)\times10^{-2}$ for $q=1$. This is still an order of magnitude lower than that for luminous blazars and gamma-ray burst having $\epsilon_{\rm rad,jet}\sim0.1$\citep{nem12,ghi14} which assume cold protons only and $q=1$. Therefore, the radiative efficiency of HBLs may be weaker than luminous AGN jet populations as the both assumptions on $\gamma'_{\rm min}$ give lower radiation efficiency than the luminous populations.

By setting $\gamma'_{\rm min}=m_p/m_e$ (Equation \ref{eq:gmin_ratio}), we have $(\eta_{\rm jet}, \xi_{\rm jet})= (3.0\times10^{-2}, 6.3\times10^{-4})$ and $(4.3\times10^{-2}, 9.2\times10^{-4})$ for $q=10$ and 1, respectively. FSRQs are known to have $\xi_{\rm jet}\sim0.1$--1 \citep[e.g.][]{cel08,ghi14} assuming cold protons only and $q=1$. Bolometric luminosities of HBLs and FSRQs are roughly different for 5 orders of magnitude. Adopting $\epsilon_{\rm rad,jet}$ and $\xi_{\rm jet}$ for the cold proton model with $q=1$ for HBLs, the mass accretion rate of FSRQs approximately needs to be $\dot{m}_{\rm acc}\sim0.03-0.3$ which is consistent with the standard disk accretion. The bolometric luminosity is proportional to $L_{\rm bol}\propto \epsilon_{\rm rad,jet}P_{\rm jet}\propto\epsilon_{\rm rad,jet}\Gamma\xi_{\rm jet}\dot{m}_{\rm acc}$. 

\begin{figure}
 \begin{center}
  \includegraphics[width=9.0cm]{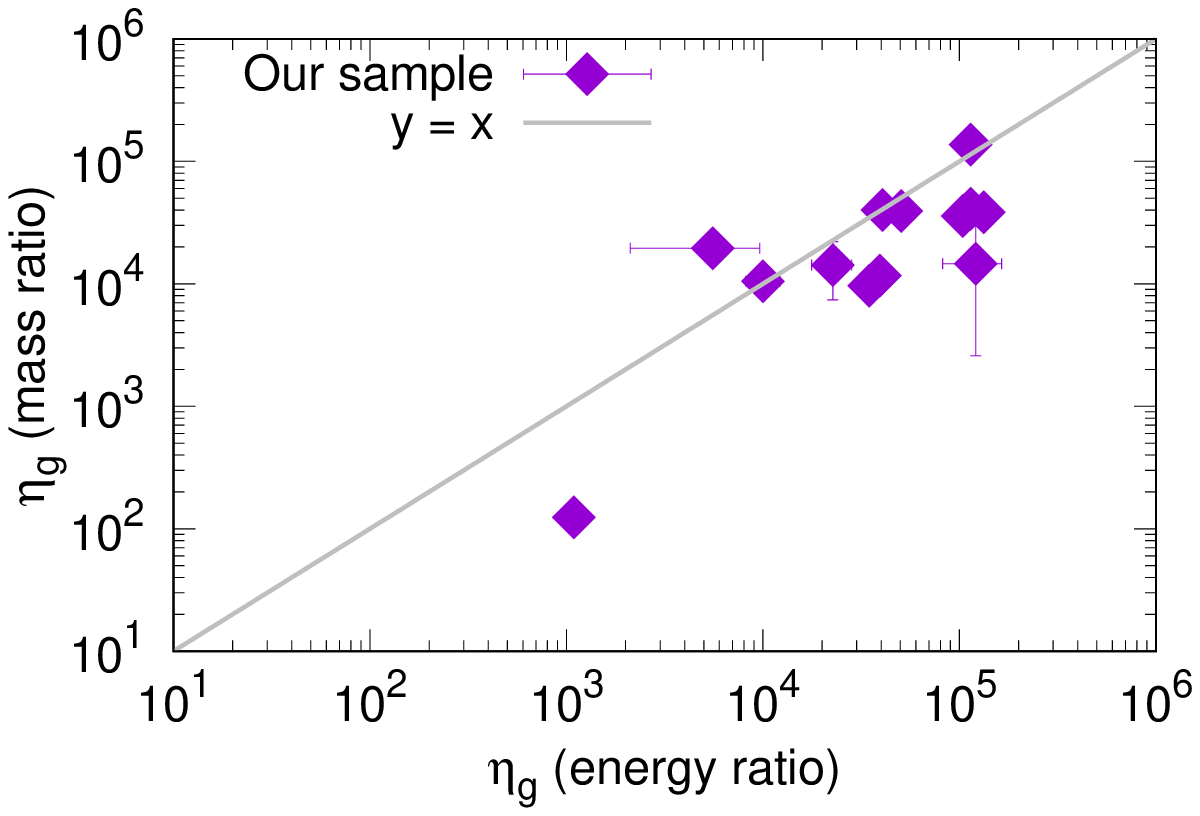} 
 \end{center}
\caption{Same as \autoref{fig:gmin_Pjet} but for $<\eta_g>$.}\label{fig:gmin_eta}
\end{figure}

\autoref{fig:gmin_eta} shows the comparison of $\eta_g$ based on models in which $\gamma'_{\rm min}$ is determined by the shocked energy  (Equation \ref{eq:gmin}) or the mass ratio of $m_p/m_e$ (Equation \ref{eq:gmin_ratio}). The expected $\eta_g$ in the both models are consistent with each other. This is because $\eta_g$ is determined by high energy X-ray synchrotron emission component (i.e. the position of the electron cutoff energy). Therefore, our result on the particle acceleration efficiency will not change even if we assume $\gamma'_{\rm min}=m_p/m_e$.

\subsection{Secondary Gamma-ray Photons from Escaped Protons}
Internal hadronic processes are known to be inefficient and requires super-Eddington jet powers $P_{\rm jet}\sim100L_{\rm Edd}$ to explain the measured photon spectra \citep{sik09,zdz15}. However, escaped high energy protons propagating intergalactic space can still generate gamma rays through cascade processes \citep[e.g.][]{ess10}. Those protons interact with the CMB and EBL photons via $p\gamma\rightarrow p\pi^0$, $n\pi^+$, and $pe^+e^-$. These interaction channels generate electromagnetic cascades distributed uniformly along the line of sight. The high energy gamma rays are produced relatively close to the Earth and not significantly affected by the EBL attenuation. The gamma-ray signals from those cascade processes are observed as point sources by current IACTs as long as the intergalactic magnetic field strength is in the femtogauss range \citep{ess11_igmf}.

The required proton power $P_p$ is $\sim10P_\gamma$ to explain the extreme blazars with secondary gamma rays, although it depends on the proton spectrum and the magnetic field structures \citep[e.g.][]{ess10, mur12}. This proton power can be affordable based on our estimate. 

In our paper, it is found that the maximum proton energy is $E_{p, {\rm max}}\sim 3\times10^{14} ({f_B}/{6\times10^{-4}})^{1/2} (P_{\rm jet}/10^{46}~{\rm erg/s})^{1/2} (\eta_g/10^{4.5})^{-1}~[\rm eV]$. Here, the required primary proton energy for secondary gamma rays is $E_p \gtrsim 10^{17}$~eV \citep[e.g.][]{ess10,mur12,tav14,yan15_2nd,zhe16}. It would be difficult to generate significant amount of secondary gamma-ray photons to explain the TeV gamma-ray data. Therefore, following the standard DSA scenario and the one-zone SSC model with the assumption that protons and electrons have the same acceleration efficiency, it may be difficult to expect secondary gamma rays from escaped protons. However, the secondary gamma-ray photons arrive with delays from $\sim$0.1--100 years for photons of $>1$~TeV for the sources at $z>0.2$ assuming proton injection energy at $10^{17}$~eV which can be longer for lower proton injection energies \citep{pro12}. Therefore, different physical parameters need to be considered for the secondary component. Moreover, as blazars are variable, TeV gamma-ray spectra can be dominated by the secondary gamma rays averaged over past faring activities. Turbulent magnetic fields can be expected at flaring states which may accelerate protons to much higher energies. Furthermore, as we discussed in Section \ref{sec:PAE}, protons can be accelerated to higher energies by large scale turbulences than we estimated.

\citet{ess12} also pointed out that sources with redshifts $z>0.15$ are more likely to exhibit the secondary gamma-ray component. In our paper, however, only 4 sources are at $z>0.15$ (Table \ref{tab:sample}). Further detailed studies of distant blazars are necessary to probe the secondary gamma-ray scenario in detail. The next generation TeV gamma-ray telescope Cherenkov Telescope Array \citep[CTA;][]{act11} will test the existence of secondary gamma-ray photons by individual distant sources \citep[e.g.][]{tak13,ess14} and by statistical samples \citep{ino14}.

\section{Summary}
\label{sec:sum}
In this paper, we study energetics and particle acceleration efficiency of 13 nearby TeV gamma-ray detected HBLs using multi-wavelength spectral analysis. We use the {\it MAXI}/GSC, {\it Swift}/BAT, {\it Fermi}/LAT, and low-state IACTs data. As leptonic SSC models successfully reproduce HBL SEDs, we consider one-zone SSC models. Assuming electrons are accelerated by the DSA processes, we are able to determine the minimum electron Lorentz factor $\gamma'_{\rm min}$, the cooling break electron Lorentz factor $\gamma'_b$, and the maximum electron Lorentz factor $\gamma'_c$ from the given parameters. The free parameters in our studies are the normalization of the electron distribution ($K_e$), the indices of the electron distribution ($p_1$ and $p_2$), the magnetic field strength ($B$), the beaming factor ($\delta$), and the gyrofactor ($\eta_g$). $\eta_g$ corresponds to the particle acceleration efficiency in the jet. By performing multi-wavelength spectral fits, we determine parameters. From the inferred parameters, we estimate physical quantities of jets such as energetics and maximum proton energies. 

We find $<p_1>=2.3$ which is in agreement with the relativistic shock acceleration theory which expect $p_1\sim2.2$ \citep[e.g.][]{kir00,ach01,kes05,sir15_review}. $\Delta p$ should be unity considering radiative cooling \citep[e.g.][]{lon94}. From our HBL samples, we find $\Delta p= 0.64$. $\Delta p=1$ leads the spectral steepening of 0.5 for synchrotron and IC emissions in the homogeneous stationary fluid. However, for inhomogeneous sources, different amounts of radiation spectral steepening can be achieved depending on geometrical structures \citep{rey09}. Therefore, the departure from $\Delta p=1$ in each source may reflect the intrinsic inhomogeneity of the emitter which is not modelled in our study. 

We find $<P_B/P_e>=6.2\times10^{-3}$ for our HBL samples. The departure from equipartition is consistent with previous studies \citep[e.g.][]{tav98,cel08,tav10,ghi10,zha12,tav15}. Such low magnetic power makes magnetic reconnections difficult to accelerate particles in HBLs \citep{sir15}. The radiative efficiency is found to be $<P_{\rm rad}/P_{\rm jet}>=6.7\times10^{-4}$, while it is known that $\epsilon_{\rm rad,jet}\sim0.1$ for luminous blazars and gamma-ray bursts \citep{nem12,ghi14}. We assume that 10\% of the shocked energy goes into lepton acceleration and the jet composition is 10 leptons per one proton, while previous studies assumed cold protons only and one electron per one proton.

In this paper, we also take into account the masses of the central SMBHs. This enables us to find the relation between the jet power and the BH mass as $<P_{\rm jet}/L_{\rm Edd}>=0.18$. Furthermore, HBLs are known to have RIAF disks \citep[e.g.][]{ghi10}. The typical mass accretion rate of HBLs is $\dot{m}_{\rm acc}\simeq1.2\times10^{-2}$ \citep{wan02}. We get the jet production efficiency $\eta_{\rm jet}\equiv P_{\rm jet}/\dot{M}_{\rm acc}c^2\sim1.5$ and $\xi_{\rm jet}\equiv\dot{M}_{\rm jet}/\dot{M}_{\rm acc}\gtrsim0.05$ under our model assumptions. However, by setting $\gamma'_{\rm min}=m_p/m_e$, those become $(\eta_{\rm jet}, \xi_{\rm jet})= (3.0\times10^{-2}, 6.3\times10^{-4})$ due to the increase of the minimum Lorentz factor.

By taking into account the latest hard X-ray spectral data, we can evaluate the particle acceleration efficiency in the blazar zone. The acceleration efficiency $\eta_g$ inferred from our spectral fits clusters at $\sim10^{4.5}$ which is consistent with previous studies of individual sources \citep{ino96,sat08,fin08}. Following the Hillas condition, we can have $E_{p,{\rm max}}\sim 3\times10^{14} ({f_B}/{6\times10^{-4}})^{1/2} (P_{\rm jet}/10^{46}~{\rm erg/s})^{1/2} (\eta_g/10^{4.5})^{-1}~[\rm eV]$. This energy is much lower than the energy of UHECRs even if we consider the uncertaities of $\eta_g$ or iron element. The blazar zone in low states can not be the UHECR acceleration sites under the DSA scenario. Such low proton maximum energy also makes HBLs difficult emit TeV--PeV neutrinos and secondary gamma rays.

\bigskip
The authors would like to thank Naoki Isobe for providing the {\it MAXI}/GSC catalog data. The authors also thank Maxim Barkov, Charles Dermer, Akihiro Doi, Yasushi Fukazawa, Susumu Inoue, Alexander Kusenko, Herman (Shiu-Hang) Lee, Greg Madejski, Jin Matsumoto, and Kohta Murase for useful comments and discussions. This research has made use of the NASA/IPAC Extragalactic Database (NED) which is operated by the Jet Propulsion Laboratory, California Institute of Technology, under contract with the National Aeronautics and Space Administration.

\vspace{5mm}
\facilities{{\it MAXI}/GSC, {\it Swift}/BAT, {\it Fermi}/LAT, H.E.S.S., MAGIC, VERITAS}

\end{document}